\documentclass[a4paper,11pt]{article}
\pdfoutput=1 
             
\usepackage{jheppub} 
                     
\pdfoutput=1
\usepackage{graphicx,array}
\usepackage{color}
\usepackage{setspace}
\usepackage{amstext}
\usepackage{amssymb}
\usepackage{slashed}
\usepackage{makecell}
\usepackage{multirow}
\usepackage{wasysym}
\usepackage{physics}
\usepackage[T1]{fontenc} 
\usepackage{changepage}
\usepackage{arydshln}
\usepackage{tikz}
\usetikzlibrary{tikzmark}           
\usepackage{scalerel} 

\definecolor{orange}{rgb}{1,0.5,0}
\definecolor{brown}{rgb}{0.59, 0.29, 0.0}

\definecolor{note_fontcolor}{rgb}{0.80078125, 0.80078125, 0.80078125}

\definecolor{darkgreen}{rgb}{0,0.5,0}

\def\beq{\begin{equation}}
\def\eeq{\end{equation}}
\def\bea{\begin{eqnarray}}
\def\eea{\end{eqnarray}}

\def\beq{\begin{equation}}
\def\eeq#1{\label{#1}\end{equation}}
\def\eeqn{\end{equation}}
\def\beqa{\begin{eqnarray}}
\def\eeqa#1{\label{#1}\end{eqnarray}}
\def\eeqan{\end{eqnarray}}
\def\CR{\nonumber \\ }
\def\leqn#1{(\ref{#1})}

\newcommand{\centeron}[2]{{\setbox0=\hbox{#1}\setbox1=\hbox{#2}\ifdim
		\wd1>\wd0\kern.5\wd1\kern-.5\wd0\fi \copy0
		\kern-.5\wd0\kern-.5\wd1\copy1\ifdim\wd0>\wd1
		\kern.5\wd0\kern-.5\wd1\fi}}
\newcommand{\ltap}{\>\centeron{\raise.35ex\hbox{$<$}}
	{\lower.65ex\hbox{$\sim$}}\>}
\newcommand{\gtap}{\>\centeron{\raise.35ex\hbox{$>$}}
	{\lower.65ex\hbox{$\sim$}}\>}

\newcommand{\lsim}{\mathrel{\ltap}}
\newcommand{\diff}{\mathrm{d}}

\def\pmu{|{\bf p}, \mu^2\rangle}

\makeatletter
\newcommand*{\Relbarfill@}{\arrowfill@\Relbar\Relbar\Relbar}
\newcommand*{\xeq}[2][]{\ext@arrow 0055\Relbarfill@{#1}{#2}}
\makeatother

\usepackage{cleveref}

\title{\boldmath Continuum Dark Matter}

\author[a]{Csaba Cs\'aki,}
\author[a,b,c]{Sungwoo Hong,}
\author[a,d]{Gowri Kurup,}
\author[e]{Seung J. Lee,}
\author[a]{Maxim Perelstein,}
\author[f]{and Wei Xue}

\affiliation[a]{Department of Physics, LEPP, Cornell University, Ithaca, NY 14853, USA}
\affiliation[b]{Department of Physics, The University of Chicago, Chicago, IL 60637 , USA }
\affiliation[c]{Argonne National Laboratory, Lemont, IL 60439, USA}
\affiliation[d]{Rudolf Peierls Centre for Theoretical Physics, University of Oxford, Parks Rd, Oxford OX1 3PJ, United Kingdom}
\affiliation[e]{Department of Physics, Korea University, Seoul, 136-713, Korea}
\affiliation[f]{Department of Physics, University of Florida, Gainesville, FL 32611, USA}

\abstract{We initiate the study of dark matter models based on a gapped continuum. Dark matter consists of a mixture of states with a continuous mass distribution, which evolves as the universe expands. We present an effective field theory describing the gapped continuum, outline the structure of the Hilbert space and show how to deal with the thermodynamics of such a system. This formalism enables us to study the cosmological evolution and phenomenology of gapped continuum DM in detail. As a concrete example, we consider a weakly-interacting continuum (WIC) model, a gapped continuum counterpart of the familiar WIMP. The DM interacts with the SM via a Z-portal. The model successfully reproduces the observed relic density, while direct detection constraints are avoided due to the effect of continuum kinematics. The model has striking observational consequences, including continuous decays of DM states throughout cosmological history, as well as cascade decays of DM states produced at colliders. We also describe how the WIC theory can arise from a local, unitary scalar QFT propagating on a five-dimensional warped background with a soft wall.}

\begin{document}
\maketitle
\flushbottom

\section{Introduction}
\label{sec:Intro}

The microscopic nature of dark matter (DM) remains one of the most important outstanding questions 
in fundamental physics~\cite{Bertone:2016nfn}. 
DM cannot consist of any of the Standard Model (SM) particles, providing firm evidence for new physics. 
Many models have been proposed by theorists, covering a mass range between $10^{-22}$ to $10^{67}$ eV, as well as various interaction portals to the SM~\cite{Bertone:2010zza,Bergstrom:2009ib,Feng:2010gw}. 
Extensive experimental efforts are underway aiming to either detect non-gravitational signatures of ambient DM, 
or to produce DM particles in the lab~\cite{gaskins2016review,Schumann:2019eaa,Kahlhoefer:2017dnp}. 
Both approaches can yield powerful hints to illuminate the nature of DM; however, so far, neither has been successful in detecting signals of DM. In fact so far we don't even know for sure whether DM consists of an elementary particle, composite bound states \cite{Frigerio:2012uc,Cline:2013zca}, extended objects such as Q-balls \cite{Kusenko:1997si}, or even macroscopic entities such as 
primordial black holes \cite{Hawking:1971ei, Carr:2016drx, Sasaki:2018dmp} and 
ultra-compact mini-halos \cite{Hogan:1988mp, Kolb:1993zz, Kolb:1993hw, Zurek:2006sy, Hardy:2016mns}.
It is therefore timely to explore the spectrum of theoretical possibilities for DM, as such explorations are both important in their own right and can provide guidance essential for future experimental searches. 
	
	In this paper, we propose a new conceptual framework in which dark matter is described by a {\it gapped continuum}, rather than an ordinary particle. In quantum field theories (QFTs) with a gapped continuum, the singly-excited states are characterized by a continuous parameter $\mu^2$, in addition to the usual 3-momentum ${\bf p}$. The parameter $\mu^2$ plays the role of mass in the kinematic relation $p^2=\mu^2$ for each state. The number of states is proportional to $ \int \rho(\mu^2)\,d\mu^2$, where $\rho$ is the {\it spectral density} of the theory, conventionally defined in QFT's as 
\bea
\langle 0 | \Phi (p)  \Phi (-p) |0 \rangle = \int \frac{d \mu^2}{2\pi} \frac{i \rho (\mu^2 )}{p^2 - \mu^2 + i \epsilon}\ .
\eea
The word ``gapped'' refers to  continuum QFTs in which the function $\rho$ has no support below some finite {\it gap scale}, $\mu_0^2$. For application to DM physics we will not be concerned with the exact origin of the continuum. We will simply assume that some form of dynamics created such a continuum as its effective description, and will treat it as a ``free continuum'', as discussed in detail in \cref{sec:physics_gapped_continuum}. Later in \cref{sec:DM_spectral_density_5D} we will provide one possible origin of such a continuum by considering a warped extra dimensional soft wall background. 

\begin{figure}
\begin{center}
\includegraphics[width=10cm]{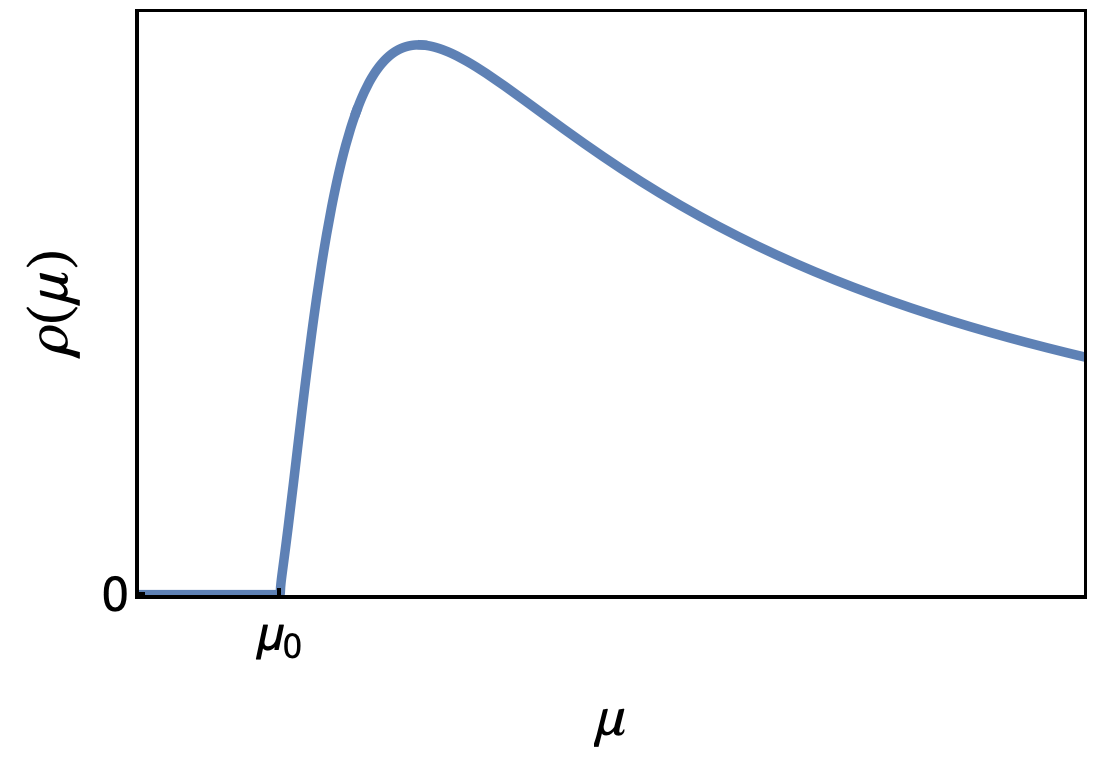} 
\end{center}
\caption{A typical shape of the spectral density $\rho (\mu)$ with gap scale $\mu_0$. }
\label{fig:rho_example} 
\end{figure}

We will construct DM models based on continuum QFT with the gap around the electroweak scale, $\mu_0 \sim 100$ GeV, and including interactions to the electroweak (EW) sector of the SM. We will call the resulting type of model the Weakly Interacting Continuum (WIC) DM model. A typical spectral density for the class of theories we consider is shown in \cref{fig:rho_example}.  

The key feature of WIC models is that DM cannot be thought of as a gas of particles of the same mass, or even as a mixture of gases of a finite number of DM species with different masses.\footnote{
In~\cite{Katz:2015zba,Chaffey:2021tmj}, a continuum is used as the mediator to the  dark sector, while here the continuum is the dark sector itself.}
 Instead, in cosmology DM states follow a continuous mass distribution determined by the product of spectral density and occupation number. In the expanding universe, this distribution is time-dependent (and can in principle depend on spatial location as well), and its evolution is governed by generalized Boltzmann equations that we will discuss. In our model, the distribution of the DM mass today is clustered in a narrow window slightly above the gap scale, but in the early universe it was much more broadly distributed.    

Continuum DM has striking phenomenological consequences. For example, while the DM state at the very bottom of 
the spectrum can be stable, {\it e.g.} due to a discrete symmetry, any other state in the continuum must be unstable 
with respect to decaying into lighter DM states. The continuum DM gas exists in a state of permanent decay. This 
leads to distinctive cosmological signatures and bounds. In particular, it is possible that DM decays can re-ionize 
the universe during the ``dark ages'', and cosmic microwave background (CMB) observations place an important bound 
on the model. Another striking feature is the absence of elastic scattering of 
DM on an SM particle such as a nucleon or an electron: in a continuum theory, any such scattering induces a change in 
the DM state's mass, and is therefore inelastic. A non-relativistic DM state with mass near the gap scale can only 
scatter into a narrow band of the continuum states due to kinematic constraints. Because of this, direct detection rates 
for continuum DM are strongly suppressed with respect to a comparable single-particle DM model. This will allow us to 
build a model in which DM communicates with the SM via a $Z$-portal, has thermal relic density, but is nevertheless not 
ruled out by direct detection experiments~\cite{Aprile:2018dbl,Cui:2017nnn,Akerib:2016vxi,Escudero:2016gzx}. Colliders, on the other hand, can provide a spectacular signature of continuum DM: a typical DM state produced in a collider would undergo multiple decays into progressively lighter DM states within the detector, with softer SM energy deposits at each step and a collider-stable DM state at the end of the cascade appearing as missing energy.    

We emphasize that many predictions of the continuum DM model, including thermal freeze-out (relic density), late decays (reionization and CMB), annihilation in the DM halos (indirect detection), and scattering with target nuclei or electrons (direct detection), are governed by the shape of the spectral density very close to the gap scale $\mu_0$. Remarkably, for a broad class of gapped continuum QFT's, $\rho (\mu)$ near the gap scale takes a universal shape of the form $\rho (\mu) \propto \sqrt{\mu^2/\mu_0^2 - 1}$. We show this in \cref{sec:DM_spectral_density_5D} and \cref{appendix:spectral_density_near_gap}. This feature makes the continuum DM physics highly model independent. On the other hand, colliders can play a complementary role in uncovering the full picture of the continuum sector.

The appearance of a continuum is very common in QFT's. The spectrum of conformal field theories (CFTs) necessarily forms a continuum since the theory does not admit any mass scales. Georgi's unparticles \cite{Georgi:2007ek, Grinstein:2008qk} also describe a continuum, and have been widely used for various particle physics applications. While continuum with a mass gap has been less commonly used, one can still find many examples of a gapped continuum both in particle and condensed matter physics. In string theory such a gapped continuum shows up when one has a large number of D3 branes distributed on a disc (which is dual to ${\cal N}=4$ SUSY broken to ${\cal N}=2$ via masses for two chiral adjoints, for a related large literature see~\cite{Gubser:2000nd, Freedman:1999gk, Kraus:1998hv}). In particle physics the simplest example of a gapped continuum was proposed by Cabrer, von Gersdorff and Quiros (CGQ) \cite{Cabrer:2009we}, based on a warped extra dimension, which is the construction we will also be relying on most in this paper. Gapped continuum has been applied to Higgs physics~\cite{Falkowski:2008fz,Stancato:2008mp,Falkowski:2008yr, Falkowski:2009uy,Bellazzini:2015cgj}, and also used for  ``continuum top partners'' \cite{Csaki:2018kxb} (see also~\cite{Megias:2019vdb,Megias:2021mgj}).  A gapped continuum also readily shows up in condensed matter physics, for example the edge modes in the quantum Hall effect, or the spectral density around a quantum critical point~\cite{Fradkin:1991nr, sachdev2007quantum}. There are also well-known examples in $d<4$ dimensions such as 2d Ising model \cite{McCoy:1978ta, McCoy:1978ix}, 2d $SU(N)$ Yang-Mills theory in large-$N$ limit \cite{Wu:1977hi}, and 2d $SU(2)$ Thirring model \cite{Luther:1976mt}.

A somewhat similar framework~\cite{Dienes:2011ja,Dienes:2011sa} called Dynamical Dark Matter (DDM) has been extensively investigated in a series of papers by Dienes, Thomas and collaborators~\cite{Dienes:2012yz,Dienes:2012cf,Dienes:2013xya,Dienes:2014via,Dienes:2014bka,Boddy:2016fds,Curtin:2018ees,Dienes:2019krh}. The main premise in DDM is to have a collection (or tower) of particles that due to their exponentially long lifetimes will form a realistic multi-component dark matter. Some of the phenomena characteristic of continuum DM and the resulting WIC models investigated here do have corresponding counterparts in DDM. There will be a modification of the direct detection cross sections and energy spectra~\cite{Dienes:2012cf,Dienes:2013xya}, novel collider phenomenology signals~\cite{Dienes:2012yz,Dienes:2014bka,Dienes:2019krh}, decays among the constituents of the dark sectors leading to an evolving DM distribution~\cite{Dienes:2014via,Curtin:2018ees}, etc. However many of the essential properties of the WIC models turn out to be quite different from generic predictions of DDM. While it might be tempting to view continuum DM simply the $\Delta m\to 0$ limit of DDM it is clear that in order to achieve that a very special construction is needed. Generic extra dimensional models usually do not produce a viable dark matter sector in the limit when a KK spectrum becomes continuous. For example, a scalar field with a bulk mass in flat 5D space will produce a gapped continuum spectrum in the limit when the size of the 5th dimension is infinite (see \cref{app:flat_X_dim}), but in this limit gravity will be five-dimensional at all scales, so the model is not phenomenologically viable. 
Another familiar example, slices of AdS space, in the limit when the KK spacing vanishes generically produce a gapless continuum, which cannot play the role of dark matter. Having both a gapped spectrum and a consistent 4D gravity at long distances is a very non-trivial requirement, satisfied by the soft-wall setup we consider here. 
One nice advantage of this setup is that it is easy to impose an exact $Z_2$ symmetry that will stabilize the dark matter against decaying into pure SM final states (while still allowing decays among the dark matter states). Without such a stabilizing symmetry, dark matter stability on cosmological time scales is not generic. One crucial aspect of the WIC models is the strong kinematic suppression of the direct detection cross section, which allows us to construct a viable Z-portal WIC model. For this it is essential that the direct detection cross section is completely dominated by inelastic scattering, and that the spectral density is dominated by the region that is kinematically inaccessible in direct detection experiments. For ordinary particles as in DDM, elastic scattering will always be allowed, and generically an enhancement of the cross section is expected rather than suppression. The evolution of the dark matter distribution at late times is a common feature of both models. This aspect is providing the strong CMB constraints on these models. Continuum DM makes a simple and universal prediction that each DM state undergoes $\sim 1$ decays in every Hubble time. The CMB bound can then be translated into a lower bound on the DM coupling to the SM, with interesting consequences for collider and other phenomenology. In DDM, the implications of the CMB bound are model-dependent. Collider physics signals are expected to have more similarities between DDM and WIC models, though we will only tangentially touch this aspect of phenomenology in this paper.

The paper is organized as follows.   
We start with a preview of the essential phenomenological features of continuum dark matter in \cref{sec:preview}.
We then describe the theoretical formalism that allows us to derive physical predictions from gapped continuum QFT's, both at zero temperature and in thermodynamics. This is the subject of \cref{sec:physics_gapped_continuum}. 
In \cref{sec:freeze-out_DM}, we specialize to DM physics and derive the Boltzmann equation (BE) for 
continuum DM freeze-out. In order to demonstrate the use of our formalism, we study the freeze-out of scalar continuum DM in a simple toy model.  
Next, in \cref{sec:Z_portal_model}, we present the fully realistic continuum $Z$-portal model and use it to calculate the relic abundance of DM. \Cref{fig:Z-portal_relic} illustrates the parameter space that can reproduce the observed relic density as well as satisfy all relevant experimental constraints. (The detailed discussion of phenomenology of the $Z$-portal continuum DM is contained in the companion paper~\cite{WIC_PRL}.) In \cref{sec:DM_spectral_density_5D}, we provide a more complete description of the continuum $Z$-portal, based on a theory in warped five-dimensional spacetime (soft-wall background). 
Finally, we present our conclusion and outlook in \cref{sec:conclusion}. Some of important topics are presented in the form of appendices. 
In \cref{app:flat_X_dim}, we study a scalar field in a flat 5D and show that its 4D spectrum can be interpreted as a gapped continuum. This exercise yields interesting insights about the interpretation of spectral density, the Hilbert space of gapped continuum theories, and their thermodynamics.  
In \cref{appendix:spectral_density_near_gap}, we provide a general proof for the properties of the spectral density near the gap scale. Finally in \cref{app:AdS_CFT}, we describe the 4D dual description of our 5D warped model. We obtain the CFT dual picture in terms of canonically normalized composite continuum modes, which mix with external/elementary fields. We show that once we resum these mixings, rates computed using the standard 4D formulae reproduce the equations introduced in \cref{sec:physics_gapped_continuum}.

\section{Preview of Continuum Effects in DM Phenomenology}
\label{sec:preview}

Before presenting a systematic discussion of the physics of a gapped continuum, this section gives a preview of the novel aspects of continuum physics that will distinguish it from ordinary particle-based DM models.

\subsection*{Late decay} 
\label{sec:latedecay}

One of the important distinguishing features of continuum DM are the decays 
\beq
{\rm DM}(\mu_1) \to {\rm DM}(\mu_2) + {\rm SM}\,, 
\eeq{decay_generic}
where DM$(\mu)$ refers to a dark matter state of mass $\mu$, while ``SM'' denotes one or more SM particles. Since all continuum states carry the same quantum numbers (including any stabilizing symmetry that prevents DM decays to fully SM final states), such decays will necessarily occur continuously throughout the history of the universe. This is in sharp contrast with particle DM models, where there is at most a handful of long-lived states decaying at specific epochs determined by their intrinsic decay widths.    

In the early universe, DM is in thermal and chemical equilibrium with the SM, and the mass distribution of the DM states is 
determined by the product of spectral density and the Boltzmann factor (for details, see 
\cref{subsec:equilibrium_thermodynamics}). As the temperature drops below the gap scale $\mu_0$, the DM decouples from the SM and the total number of DM states is frozen out, just like for the usual thermal-relic particle DM. However, the mass distribution 
of the DM states continues to evolve, thanks to decays~\leqn{decay_generic}. 
The decays shift the distribution towards lower masses, closer to the gap scale. Lifetime of a DM state $\Gamma^{-1}(\mu)$ increases with decreasing mass, due to both phase-space suppression and the fact that there are fewer states for it to decay into. Schematically,
\beq
\Gamma(\mu) \sim g^2 \Delta^p \mu_0^{1-p}\,,
\eeq{gamma_generic}
where $g$ is the strength of DM-SM coupling, $p$ is a model-dependent positive number, and $\Delta=\mu-\mu_0$. If $t_0\sim H^{-1}$ is the age of the universe, only DM states for which $\Gamma \lsim H$ can still be present. This condition can be used to find the typical mass of the DM states at any given time. For example, in the model considered in detail in this paper, the DM states are currently clustered within a few hundred keV above the gap scale. It also indicates that on average, each DM state undergoes roughly one decay per Hubble time, or in other words an order-one fraction of DM states will decay during each doubling of the scale factor.   

The continuous DM decays also lead to potentially observable effects in cosmology. If the SM particles produced in the decay interact electromagnetically ({\it i.e.} all SM particles except neutrinos), the decays that occur after CMB decoupling can reionize hydrogen, drastically changing the optical depth for CMB photons \cite{Chen:2003gz, Pierpaoli:2003rz, Slatyer:2012yq, Slatyer:2015jla}. This places a stringent bound on the structure of the continuum DM models. In the Z-portal model considered in this paper, the bound can be satisfied only if the DM decays to electron-positron pairs are kinematically forbidden, $\Delta \lsim$ MeV, at and after the CMB decoupling time. This condition implies a {\it lower} bound on the strength of the DM coupling to SM, see \cref{fig:Z-portal_relic}.

\subsection*{Direct detection} 
\label{sec:DD}

A very important and generic feature of gapped continuum DM is the suppression of direct detection rates. The scattering process relevant for direct detection is
\beq
{\rm DM}(\mu_1) + {\rm SM} \to {\rm DM}(\mu_2) + {\rm SM}.
\eeq{DD_process}
The cross section can be schematically written as
\bea
\sigma \sim \int \frac{d \mu_2^2}{2\pi} \rho (\mu_2^2) \; \hat{\sigma} \left( \mu_1, \mu_2 \right),
\label{eq:rate_continuum_1}
\eea
where $\hat{\sigma}$ is an ordinary particle $2\to 2$ cross section, with the masses for the external particles replaced by the continuum parameters $\mu_1$ and $\mu_2$. If the incoming DM state has mass $\mu_1=\mu_0+\Delta$, the range of kinematically accessible values of $\mu_2$ is $[ \mu_0, \mu_0 + \Delta + Q]$, where $Q$ is the kinetic energy of the collision in the center-of-mass frame. As we argued above, continuous DM decays generically result in $\Delta\ll \mu_0$ in today's universe, while $Q\ll \mu_0$ as long as ambient DM is non-relativistic. We can then estimate 
\bea
\sigma_{\rm cont} \sim  \left( \frac{\Delta+Q}{\mu_0}\right)^{1+r}\,\sigma_{\rm particle},
\label{eq:rate_continuum_1}
\eea
where $r$ is a positive number that depends on the behavior of the spectral density near the gap. (It will be shown in \cref{sec:DM_spectral_density_5D} and \cref{appendix:spectral_density_near_gap} that $r=1/2$ in a broad class of models of gapped continuum.) For example, in the specific model that will be considered in detail in this paper, continuous DM decays result in $\Delta\sim 100$ keV at the present time, while $Q\sim 1$ keV in the case of ambient weak-scale DM colliding with a nucleus. With $\mu_0$ at the weak scale, this mechanism gives a spectacular suppression of the direct detection cross section by several orders of magnitude compared to a particle model with the same mass scale and interaction strength. We emphasize that this effect is entirely due to the continuous nature of the DM spectrum: intuitively, the suppression arises because only a tiny fraction of the DM spectrum is kinematically accessible in the scattering process~\leqn{DD_process} in a direct detection experiment.  

In contrast, indirect detection relies on annihilation processes of the form ${\rm DM}(\mu_1) + {\rm DM}(\mu_2)\to {\rm SM_1}  + {\rm SM_2}$. Since there is no continuum state in the final state, the rates of these processes are unsuppressed. In fact, since $\mu_1\approx \mu_2\approx \mu_0$ in the current universe, both rates and kinematics of annihilation in the galactic halos are virtually identical for continuum and particle DM.

\subsection*{Colliders} 
\label{sec:colliders}

\begin{figure}
	\begin{center}
		\includegraphics[width=12cm]{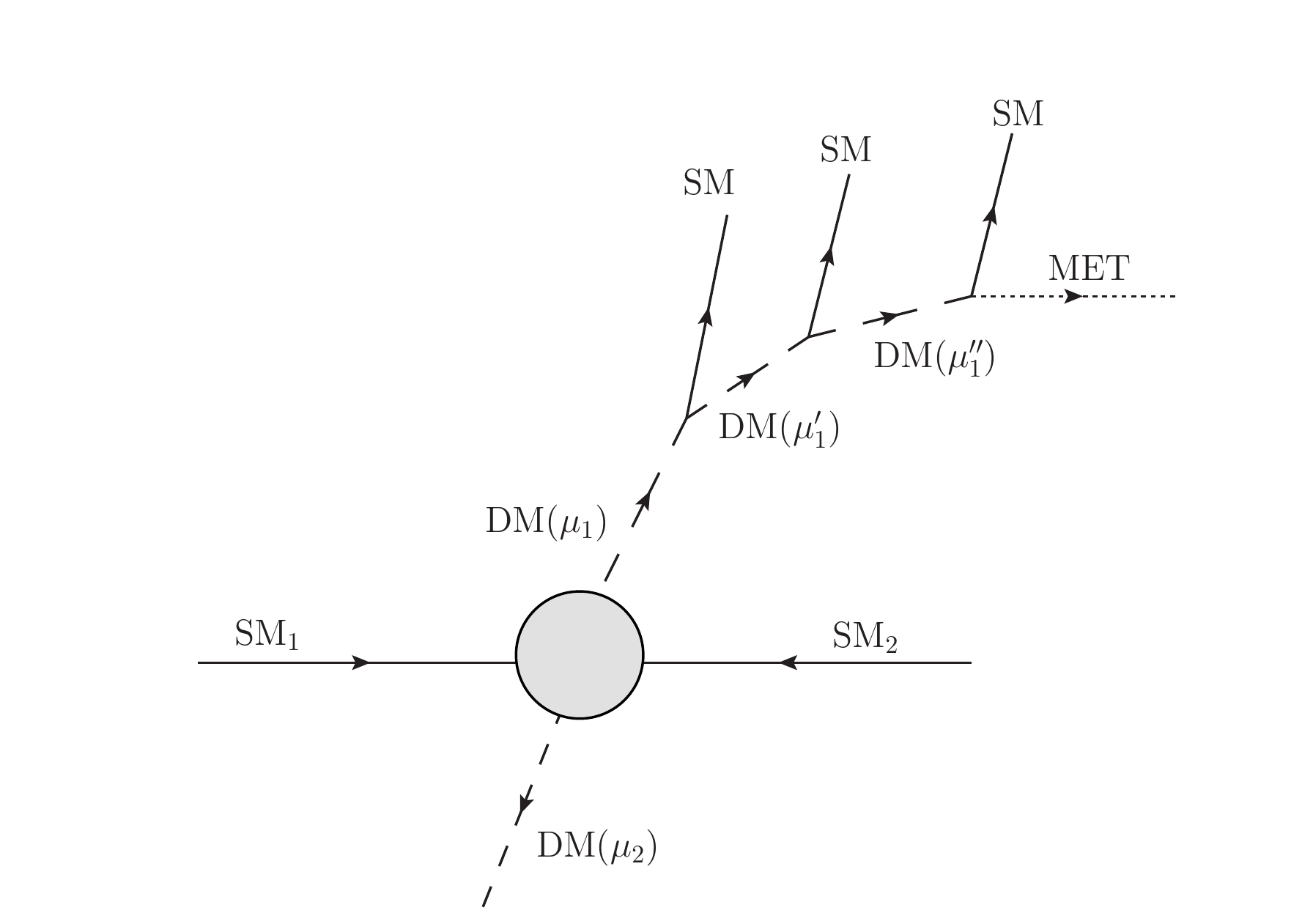} 
	\end{center}
	\caption{Schematic representation of collider signature of the continuum DM. }
	\label{fig:collider} 
\end{figure}

Continuum DM states can be produced in colliders via 
\beq
{\rm SM_1}  + {\rm SM_2}\to {\rm DM}(\mu_1) + {\rm DM}(\mu_2).
\eeq{COLL_process}
All kinematically accessible DM modes will generically be produced. The total production cross section is schematically 
\bea
\sigma \sim \int \frac{d \mu_1^2}{2\pi} \rho (\mu_1^2) \;\int \frac{d \mu_2^2}{2\pi} \rho (\mu_2^2)\; \hat{\sigma} \left( \mu_1, \mu_2 \right).
\label{eq:rate_continuum_collider}
\eea
If the collision energy is close to the threshold, the continuum kinematics leads to suppressed rates, similar to the case of direct detection. This effect may weaken collider bounds on the model. On the other hand, if collision energy is above the gap scale by an order-one factor, there is no kinematic suppression factor as in the case of direct detection, and the collider cross section is of the same order for continuum and particle DM. However, unlike particle DM, the continuum DM states quickly decay, see eq.~\leqn{decay_generic}. In fact, each state undergoes a series of decays, illustrated in \cref{fig:collider}. Each decay produces SM particles (with progressively smaller energies at each step of the cascade) in addition to a DM state (with mass closer to the gap scale at each step of the cascade). Many such decays will occur within the detector, resulting in a high-multiplicity observable SM final state with characteristic pattern of energy distributions, in addition to missing energy due to the escaping long-lived DM states. A detailed study of this exciting and novel collider phenomenology will be pursued in future work. (Similar cascade decay signatures in Dynamical Dark Matter models have been studied in Refs.~\cite{Dienes:2012yz,Dienes:2014bka,Dienes:2019krh}.)

\section{Physics of Gapped Continuum}
\label{sec:physics_gapped_continuum}

It is often stated that CFT's and theories with continuum spectra do not have a particle interpretation and no S-matrix can be defined. The main reason behind this is that the interactions leading to a non-trivial fixed point are also essential for producing the continuum spectrum of the theory. If one turns off the interactions, the spectrum changes from continuum into that of an ordinary free particle, hence the asymptotic states defined in the usual manner would not capture the physics of the system properly. This however does not imply that there would be anything wrong with these theories, nor that they could not be successfully used in particle physics for BSM sectors, but rather that one needs to find an alternative approach for defining scattering processes. Instead of relying on the definition of asymptotic states obtained by turning off the interactions, we will assume that the effects of the strong interactions can be captured by the fact that there is a non-trivial continuum (with a mass gap), and described by a (potentially non-local) effective Lagrangian 
\begin{equation}
S = \int \frac{d^4 p}{(2\pi)^4} \; \Phi^\dagger (p) \Sigma (p^2) \Phi (p)
\label{eq:action_scalar_continuum}
\end{equation}
which is designed to properly reproduce the two-point function of theory
\begin{equation}
\int d^4x \; e^{ip (x-y)} \langle 0 \vert T \Phi (x) \Phi^\dagger (y) \vert 0 \rangle =  \langle 0| \Phi (p) \Phi^\dagger (-p)|0 \rangle = \frac{i}{\Sigma (p^2)} = \int \frac{d \mu^2}{2\pi} \; \frac{i \; \rho (\mu^2)}{p^2 - \mu^2 + i \epsilon},
\label{eq:scalar2ptspectralrep}
\end{equation}
where $\rho (\mu^2)$ is the spectral density. We will assume that the effective description in \cref{eq:action_scalar_continuum} is weakly coupled, hence $\Phi$ corresponding to a  ``generalized free field''~\cite{Greenberg:1961mr}\footnote{For more recent discussions on generalized free fields, see, for example,~\cite{Dymarsky:2014zja} and references therein.}. Essentially we are assuming that the resulting continuum is free, hence we will refer to this scenario as a  ``free continuum theory''. In addition we perturb around generalized free continuum by introducing additional weak couplings to $\Phi$ and assume that the underlying structure described by the spectral density remains unchanged, resulting in a weakly interacting continuum. This picture will be supported by the concrete extra dimensional construction that we introduce in \cref{sec:DM_spectral_density_5D}. $\Phi$ will be the boundary value of a bulk scalar field propagating in a non-trivial ``soft-wall''-type background, which itself is supposed to be the 5D dual of a strongly interacting 4D CFT-like theory (see \cref{app:AdS_CFT} for details). $\Sigma$ will be the brane-to-brane propagator which can be calculated for a fixed background, and we will be adding interactions of $\Phi$ with SM fields assumed to be localized on the brane.

For a given $\Sigma$ the spectral density $\rho$ can be obtained as 
\bea
\rho (p^2) = -2 \; {\rm Im} \frac{1}{\Sigma (p^2)}.
\label{eq:rho_scalar}
\eea
If $\Sigma (p^2) = \left( p^2 - m^2 +i\epsilon \right)$, this theory merely describes a free scalar \emph{particle} with mass $m^2$, corresponding to $\rho (\mu^2) \propto \delta \left( \mu^2 - m^2 \right)$.
When $\rho (\mu^2)$ has a non-vanishing support on continuum domain in $\mu^2$, the theory describes a continuum.
In particular, if the spectral density has a continuous distribution starting at $\mu^2 \geq \mu_0^2 >0$, we have a theory of gapped continuum, with gap scale $\mu_0$. Note that a continuum contribution is always present even in the case of ordinary particles when one is considering the (loop-induced) multi-particle contributions to the spectral density. In the free continuum theories we are considering there is no one-particle pole and instead we have ``tree-level'' continuum present in $\rho (\mu^2)$, which represents the intrinsic continuum states.
The best way to think of the continuum states is to view them as smeared out particles with a density of states:  a finite physical effect is obtained only as collective effect after integrating over a finite energy interval weighted by the density of states. This density of states will be identified with the spectral density $\rho (\mu^2)$. Below we will systematically build up the formalism needed to most efficiently deal with such weakly coupled continuum states originating from a generalized free field, and present the formulae that are analog of those in ordinary particle physics. We will show that in spite of the inherent differences one can  find a simple modification of the Hilbert-space construction for the free continuum that closely parallels that of ordinary particles, which will make the calculation of reaction rates quite straightforward.

 An interaction between the gapped continuum and SM fields can be introduced in a standard QFT way: just use the corresponding field $\Phi(x)$ to build a local (gauge invariant) interaction term.  
For example, the ``Higgs-portal'' interaction will be 
\begin{equation}
S_{\rm int} = \int d^4 x \; \lambda H^\dagger H (x) \Phi^\dagger \Phi (x)
\label{eq:Higgs_portal}
\end{equation}
where $H$ is the SM Higgs doublet, and $\Phi$ is the operator responsible for the gapped continuum. While this simple interaction turns out to not lead to a phenomenologically viable DM model, we will use it as a toy model to illustrate the formalism, before presenting the fully realistic model in \cref{sec:Z_portal_model} and~\cref{sec:DM_spectral_density_5D}.

\subsection{Free Continuum QFT with a Gap}
\label{subsec:gapped_continuum_QFT}

Next we will present the basic construction of a free continuum. A pedagogical introduction to these states can also be found in \cref{app:flat_X_dim} where we show how the Kaluza-Klein (KK) states in a simple flat extra dimension can be interpreted as a continuum in 4D, which will also clarify what the right completeness and orthonormality conditions should be. A quick read of \cref{app:flat_X_dim} is highly recommended before moving on here.

We will start with the description of the Hilbert space. The single-mode sector will contain (in addition to ordinary 1-particle states corresponding to the SM) additional states labeled by  $\pmu$, which are eigenstates of Hamiltonian $\hat{H}$ and 3-momentum $\hat{\bf P}$ such that
	\beqa
	\hat{\bf P}\,\pmu &=& {\bf p} \, \pmu, \CR
	\hat{H}\,\pmu &=& E_\mu \, \pmu, \;\;\;\;\; E_\mu \equiv \sqrt{ {\bf p}^2+\mu^2}.
	\eeqa{states}
These states form the free continuum, parametrized by the continuous parameter $\mu^2$. The spectral density $\rho(\mu^2)$ can then be interpreted as the density of states with respect to this parameter. One can also introduce creation operators $a^\dagger_{{\bf p},\mu}$ for the free continuum such that 
\begin{equation}
\pmu = \sqrt{\frac{ 2E_\mu}{\rho (\mu )}} \; a^\dagger_{{\bf p},\mu} |0\rangle , \ \ \  \left[ a_{{\bf p},\mu}, a^\dagger_{{\bf p'},\mu'} \right] = (2\pi)^4 \delta^3 ({\bf p}-{\bf p'}) \delta (\mu^2 - \mu^{\prime 2}) 
\label{eq:continuum_state_n_ETC}
\end{equation}
which can also be used for the usual decomposition for the field $\Phi$ of the QFT\footnote{For complex scalar $\Phi$, we instead have
\bea
\Phi (x) = \int \frac{d\mu^2}{2\pi} \sqrt{ \rho (\mu )} \int \frac{d^3p}{(2\pi)^3 \sqrt{ 2 E_\mu }} \left( a^\dagger_{{\bf p},\mu} e^{i p\cdot x} +b_{{\bf p},\mu} e^{-i p\cdot x} \right)_{p^0=E_\mu}
\eea
with 
\bea
\left[ a_{{\bf p},\mu}, a^\dagger_{{\bf p'},\mu'} \right] = (2\pi)^4 \delta^3 ({\bf p}-{\bf p'}) \delta (\mu^2 - \mu^{\prime 2}) = \left[ b_{{\bf p},\mu}, b^\dagger_{{\bf p'},\mu'} \right] 
\eea
being the only non-trivial commutation relations.
 }
\begin{equation}
\Phi (x) = \int \frac{d\mu^2}{2\pi} \sqrt{ \rho (\mu )} \int \frac{d^3p}{(2\pi)^3 \sqrt{ 2 E_\mu }} \left( a^\dagger_{{\bf p},\mu} e^{i p\cdot x} +a_{{\bf p},\mu} e^{-i p\cdot x} \right)_{p^0=E_\mu}
\label{eq:continuum_Phi}
\end{equation}
resulting in \cref{eq:scalar2ptspectralrep}. One important subtlety regarding the free continuum theory is that the canonical momentum $\Pi (\vec{x},t)$ can differ significantly from $\dot{\Phi}$, since the Lagrangian in eq.~(\ref{eq:action_scalar_continuum}) corresponds to a higher derivative theory. Hence canonical quantization in terms of the field operator $\Phi$ becomes quite involved.\footnote{We thank Michele Redi for several useful discussions on this point.} We will instead use a holographic interpretation of $\rho (\mu^2)$ in section~\ref{subsec:5Dspectral} to fix the overall normalization of $\rho$.

%

The free continuum satisfies a completeness relation:
\beq
\int \frac{d\mu^2}{2\pi}\,\rho(\mu^2)\, \int \frac{d^3 p}{(2\pi)^3} \frac{1}{2E_\mu}\,\pmu \langle {\bf p}, \mu^2| \,=\, 1.
\eeq{compl1}
This is basically the standard one-particle completeness relation integrated over $\mu^2$  weighted by $\rho (\mu^2)$, solidifying the picture of the entire continuum effectively acting as a single ordinary particle. 
The completeness relation can also be rewritten in a nice Lorentz-invariant form 
\beq
\int \frac{d^4 p}{(2\pi)^4} \rho(p^2) \pmu \langle {\bf p}, \mu^2| \,=\, 1,
\eeq{compl2}
where $p_0 = E_\mu = \sqrt{ {\bf p}^2+\mu^2}$, and $p^2=p_0^2-{\bf p}^2$. The normalization of single-mode states consistent with completeness relation is given by
\beq
\langle {\bf p}^\prime, \mu^{\prime 2} \pmu \,=\, \frac{2E_\mu}{\rho(\mu^2)}\,(2\pi)^4 \delta^3({\bf p}-{\bf p}^\prime) \, \delta(\mu^2-\mu^{\prime 2}).
\eeq{normal}
In \cref{app:flat_X_dim}, we show that a gapped continuum can be obtained from a flat 5D space. In that simple example, completeness relation and associated state normalization are inherited from the standard 5D field theory, which indeed agree with \cref{compl1}-\cref{normal}.

Multi-mode states are built as direct products of these single-mode states, as usual. An interaction may be introduced to couple these states to the SM, and matrix elements are computed by the usual rules of perturbative QFT. For example, for a theory with \cref{eq:action_scalar_continuum} and \cref{eq:Higgs_portal}, a scattering process SM+SM$\to \Phi (\mu_1) + \Phi (\mu_2)$ is described by a matrix element
\beq
\langle ({\bf p_1}, \mu_1^2), ({\bf p_2}, \mu_2^2)| {\rm T \, exp}\left(-i \int dt H_I(t)\right) |{\bf k}_A, {\bf k}_B\rangle_{\rm SM} \equiv (2\pi)^4 \delta^4(k_1+k_2-p_1-p_2)\,i {\cal M}. 
\eeq{mel}
A measurable cross section for this process will involve the production of the continuum over a finite region of the parameter $\mu$, defined as 
\beq
\sigma = \frac{1}{2E_A}\,\frac{1}{2E_B}\,\frac{1}{|v_A-v_B|}\,\int \frac{d\mu_1^2}{2\pi}\,\rho(\mu_1^2)\,\int \frac{d\mu_2^2}{2\pi}\,\rho(\mu_2^2)\, \int d\Pi_{\mu_1}\,d\Pi_{\mu_2}\,(2\pi)^4 \delta^4(k_1+k_2-p_1-p_2)\,|{\cal M}|^2\,,
\eeq{eq:xsec}
where the Lorentz-invariant phase space (LIPS) volume element is given as usual by
\beq
d\Pi_{\mu} \,=\, \frac{d^3 p}{(2\pi)^3}\,\frac{1}{2E_\mu}.
\eeq{Pies}
In \cref{app:AdS_CFT}, we present a derivation of \cref{eq:xsec} using a warped 5D model for a continuum and AdS/CFT correspondence. 

The discussion above shows that in many respects the free continuum states are just like ordinary particles with mass $\mu^2$. The main difference is that the contribution of any single continuum state $\pmu$ to any physical process will be negligible, and only the collective effect of the continuum will give finite contributions. Hence as stated above it is best to think of the continuum as a single smeared out particle. If all continuum states are accessible in a given scattering process, their contribution is similar to that of a single ordinary particle. However, if only a fraction of continuum states are kinematically accessible, the continuum will act as a ``partial'' particle, leading to suppressed scattering: DM direct detection offers a phenomenologically relevant example of this phenomenon.  

\subsection{Equilibrium Thermodynamics}
\label{subsec:equilibrium_thermodynamics}

Next we consider a dilute, weakly-coupled, spatially uniform gas made out of the free continuum states described above. We define the dimensionless phase-space density $f({\bf p}, \mu^2)$ such that the number of excitations with mass-squared between $\mu^2$ and $\mu^2+d\mu^2$ is given by
\beq
dN = V g  \frac{d\mu^2}{2\pi}\,\rho(\mu^2)\, \int \frac{d^3 p}{(2\pi)^3} \, f({\bf p}, \mu^2),
\eeq{Ngen}
where $g$ is the number of internal degrees of freedom and $V$ is the volume occupied by the gas. (We will set $g = 1$ in the rest of this section to simplify the expressions.) 
The energy of the gas is given by 
\begin{equation}
E = V \int \frac{d\mu^2}{2\pi}\,\rho(\mu^2)\, \int \frac{d^3 p}{(2\pi)^3} \, f({\bf p}, \mu^2) E_\mu.
\label{eq:Egen}
\end{equation}
If interactions among continuum modes in the gas (either directly with each other, or through their interactions with some other, e.g.~SM, gas) are strong enough to maintain them in thermal and chemical equilibrium with each other, the state of the gas can be completely characterized by two parameters, temperature $T=1/\beta$ and chemical potential which we denote by $\eta$ (to avoid confusion with the $\mu$ parametrizing the continuum). In this case, the phase-space density takes the standard Fermi-Dirac or Bose-Einstein form:\footnote{This can be proven by the standard method: we extremize the entropy with the constraints of energy and number density \cref{Ngen} and \cref{eq:Egen}. These constraints can be enforced by means of Lagrange multiplier, and we need to extremize 
\bea
\tilde{S} = S + \beta \left( \int \frac{d \mu^2}{2\pi} \rho \int \frac{d^3p}{(2\pi)^3} f E - u \right) + \gamma \left( \int \frac{d \mu^2}{2\pi} \rho \int \frac{d^3p}{(2\pi)^3} f - n \right).
\eea
The solution to $\frac{\delta \tilde{S}}{\delta f} = 0$ is 
\bea
f^{\rm eq}  = e^{-(\gamma+1)} e^{-\beta E}
\eea
where the prefactor $e^{-(\gamma+1)}$ is fixed by the normalization.
}
\beq
f({\bf p}, \mu^2) = \frac{1}{e^{\beta(E_\mu-\eta)} \pm 1} \,\approx\, e^{-\beta(E_\mu-\eta)}  
\eeq{f_equil}  
where the two signs correspond to fermionic ($+$) and bosonic ($-$) free continuum states, and the second (Boltzmann) form applies in the limit of small occupation numbers. We will assume this limit below, and consider the case of zero chemical potential $\eta=0$ as an example. 

The free energy is given by
\beq
F = \frac{1}{\beta}\, V\, \int \frac{d\mu^2}{2\pi}\,\rho(\mu^2)\, \int \frac{d^3 p}{(2\pi)^3}\, \ln \left( 1 \pm e^{-\beta E_\mu}\right). 
\eeq{F} 
Energy density $u$ and pressure $P$ can be found through the standard thermodynamic relations, 
\beqa
u &=& \frac{1}{V}\left( \left.\beta \frac{\partial F}{\partial \beta}\right|_V \,+\,F\right)\,,\CR
P &=& \left.- \frac{\partial F}{\partial V}\right|_\beta, 
\eeqa{rhoP}
and their explicit expressions are given by
\beqa
u &=&  \int \frac{d\mu^2}{2\pi}\,\rho(\mu^2)\, {\cal U}(\mu^2)\, , \;\;\;\;\; {\cal U} (\mu^2) = \int \frac{d^3 p}{(2\pi)^3} \frac{E_\mu}{e^{\beta E_\mu} \pm 1} ,\CR
P &=&  \int \frac{d\mu^2}{2\pi}\,\rho(\mu^2)\, {\cal P}(\mu^2)\,, \;\;\;\;\;  {\cal P}(\mu^2) =  \int \frac{d^3 p}{(2\pi)^3} \frac{1}{e^{\beta E_\mu} \pm 1} \frac{p^2}{3E_\mu} 
\eeqa{rhoP_int}
where ${\cal U}$ and ${\cal P}$ are the equilibrium energy density and pressure of a gas of ``normal'' particles with mass-squared $\mu^2$. Roughly speaking, at temperatures above the gap scale, $T>\mu_0$, energy and pressure are dominated by modes with $\mu_0<\mu<T$, which behave as a relativistic gas. At temperatures below the gap scale, $T<\mu_0$, energy and pressure are dominated by modes with $\mu \approx \mu_0$ (with details depending on the behavior of the spectral density in that region), which behave as a gas of non-relativistic particles. In this regime, the continuum gas can play the role of cold dark matter (CDM).           

In principle, the spectral density itself can also be temperature dependent: $\rho (\mu^2, T)$. However, we expect that the thermal corrections will be of the order $\mathcal{O} (T/\Lambda)$, where $\Lambda\gg \mu_0$ is the cutoff scale of our description above which gapped continuum phase is replaced by a UV phase such as CFT. For this reason, in the following we will ignore this dependence, and revisit the validity of this assumption in \cref{sec:DM_spectral_density_5D} when we discuss the concrete warped 5D model of gapped continuum. 

\subsection{Non-equilibrium Thermodynamics}
\label{subsec:non_equili_thermody}

We continue to consider a dilute, weakly-coupled gas of continuum states, but now do not assume that it is in thermal and/or chemical equilibrium. In this case the phase-space density is still sufficient to describe the gas, but it can now be a function of time: $f({\bf p}, \mu^2, t)$. The time evolution of this quantity is described by the Boltzmann equation. For example, consider a model in which the continuum states can interact with SM states $A,B$ through $2\leftrightarrow 2$ scattering. In this case, the Boltzmann equation (in flat-space background) reads
\beqa
E_\mu \frac{\partial f({\bf p}, \mu^2, t)}{\partial t} &=& - \frac{1}{2} \int \frac{d\mu^{\prime 2}}{2\pi}\,\rho(\mu^{\prime 2})\, \int d\Pi_{\mu^{\prime}} \, d\Pi_A d\Pi_B \,
 (2\pi)^4 \delta^4 (k_A+k_B-p-p^\prime)\, \CR\CR & &\times |{\cal M}|^2\,\left( f f^\prime (1\pm f_A) (1\pm f_B) - f_A f_B (1\pm f)(1\pm f^\prime)\right),
\eeqa{BE}   
where the usual sums over spin are included in $\vert \mathcal{M} \vert^2$.
In the collision term on the right-hand side, $k_A$ and $k_B$ are the 4-momenta of the SM particles, $p$ and $p^\prime$ are the 4-momenta of the continuum states (note that $p^2=\mu^2$ and $p^{\prime 2}=\mu^{\prime 2}$), $d\Pi$ are the LIPS volume elements defined in eq.~\leqn{Pies}, and ${\cal M}$ is the scattering amplitude defined in eq.~\leqn{mel}. In the limit of low occupation numbers which we will consider from now on, terms with $\pm$ in front can be ignored. Generalization to gas in FRW background is straightforward. The only change is on the left-hand side, where the derivative $\partial/\partial t$ needs to be replaced with the covariant version, giving
\beqa
 E_\mu \frac{\partial f(E, \mu^2, t)}{\partial t} - & & H |{\bf p}|^2 \frac{\partial{f(E, \mu^2, t)}}{\partial E} \,=\, - \frac{1}{2} \int \frac{d\mu^{\prime 2}}{2\pi}\,\rho(\mu^{\prime 2})\, \int d\Pi_{\mu^{\prime}} \, d\Pi_A d\Pi_B \, \CR\CR & & \times 
(2\pi)^4 \delta^4 (k_A+k_B-p-p^\prime)\, |{\cal M}|^2\,\left( f f^\prime - f_A f_B\right).
\eeqa{BE_FRW}
Here $H=\dot{a}/a$ is the Hubble parameter, $|{\bf p}|^2=E^2-\mu^2$, and we replaced ${p}$ with $E$ as the argument of $f$ since 3D rotational invariance guarantees that $f$ only depends on the magnitude of ${\bf p}$. Note that if the continuum originates from a 5D model, then using this form of the Boltzmann equation implies the assumption that the geometry of the bulk is fixed and only the 4D scale factor is still evolving.

\section{Freeze-Out of Continuum Dark Matter}
\label{sec:freeze-out_DM}

As an application of the above formalism, we study the process of freeze-out of continuum DM which can interact via $2\leftrightarrow 2$ scattering with an SM particle with mass $m_{\rm SM} \ll \mu_0$ (we set $m_{\rm SM}=0$ below).
Before we give a more technical discussion, however, it seems instructive to note the following. Similarly to the particle DM, annihilation of continuum DM freezes out at $T \sim \frac{\mu_0}{10}$. At such low temperature, the continuum mass distribution is localized close to the gap scale, and it behaves more or less like a particle with mass $\sim \mu_0$. Therefore, as far as thermal freeze-out is concerned, the continuum DM is expected to be similar to particle DM. Below we confirm this by explicit computations and estimate the size of ``continuum effects''.

\subsection{Boltzmann Equation for Continuum Freeze-Out}
\label{eq:Boltzmann}

For thermal freeze-out, there are two relevant reactions, \emph{annihilation}
\beq
{\rm ~~~~~~~DM (\mu) + DM (\mu')} \leftrightarrow {\rm SM+SM}
\eeq{an}
and \emph{quasi-elastic scattering} (QES)
\beq
{\rm ~~~~~~~DM (\mu) + SM} \leftrightarrow {\rm DM (\mu') + SM}.
\eeq{el}
If annihilation is in equilibrium, the continuum DM modes are at the same temperature $T$ as the SM and  at zero chemical potential. This will be the case, for sufficiently strong coupling between the SM and DM, at temperatures above the gap scale.\footnote{For any $T>\mu_0$, some of the DM states will be non-relativistic, since one can always go sufficiently far out on the tail of the continuum DM spectral density to ensure $\mu>T$. The equilibrium density of these states is exponentially suppressed, however they still remain in equilibrium and do not freeze out, since they can find annihilation partners among lighter DM states with $\mu<T$, whose equilibrium density is unsuppressed. 
Decay and inverse decay, ${\rm DM (\mu)} \leftrightarrow {\rm DM (\mu')} + {\rm SM}$, also maintain equilibrium among the modes with different $\mu$.}
Once $T\lsim \mu_0$, however, the annihilation rate drops exponentially, and annihilations decouple (``freeze-out''). Note that the rate of quasi-elastic scattering of a DM state does not experience an exponential drop at these temperatures, and therefore the QES process continues to maintain thermal equilibrium between the SM and DM. It also maintains DM states of different masses in chemical equilibrium with each other, since the DM mass changes during QES. Therefore, during the freeze-out process, the DM modes are at the same temperature as the SM, $T$, and have a common ($\mu$-independent) chemical potential $\eta$, which however is time-dependent and no longer vanishes. This means that in the freeze-out calculations we can assume
\beq
f_{\rm DM} = e^{-\beta\left(E_\mu-\eta (t)\right)},~~~~~~f_{\rm SM} = e^{-\beta |{\rm p}|}.
\eeq{fs}    
The effective DM number density is given by
\beq
n = \int \frac{d\mu^2}{2\pi}\,\rho(\mu^2)\,\int \frac {d^3 p}{(2\pi)^3}\,f_{\rm DM}.
\eeq{nDM} 
If $n_{\rm eq}$ denotes the value of $n$ with $\eta=0$ (i.e. in chemical equilibrium with the SM), then $n=n_{\rm eq} e^{\beta \eta}$, leading to a useful expression of the DM phase space density in terms of the number densities 
\beq
f_{\rm DM} = \frac{n}{n_{\rm eq}}\,e^{-\beta E_\mu}.
\eeq{f_ratio}
Integrating both sides of the Boltzmann equation eq.~\leqn{BE_FRW} with respect to $\int \frac{d\mu^2}{2\pi}\,\rho(\mu^2)\,\int \frac {d^3 p}{(2\pi)^3}$, and using eq.~\leqn{f_ratio} on the right-hand side and the usual integration-by-parts trick in the second term on the left-hand side, the equation for time evolution of DM number density becomes, 
\beq
\frac{\partial n}{\partial t} + 3Hn = - \langle \sigma v \rangle (n^2 - n_{\rm eq}^2)\,,
\eeq{n_evol}
where we defined
\begin{align}
\langle \sigma v \rangle &= \frac{1}{n_{\rm eq}^2}\, \int \frac{d\mu^2}{2\pi}\,\rho(\mu^2)\int \frac{d\mu^{\prime 2}}{2\pi}\,\rho(\mu^{\prime 2}) \, \int d\Pi_\mu \, d\Pi_{\mu^{\prime}} \, d\Pi_A d\Pi_B \,\CR & \times (2\pi)^4 \delta^4 (k_A+k_B-p-p^\prime)\, |{\cal M}|^2\, \exp\left(-\beta(E_A+E_B)\right)\,.
\label{eq:sigma_v}
\end{align}

It is interesting to note that \cref{n_evol} is identical to that of the usual particle cold relic, and hence the relic density is given by the usual expression found in e.g. Kolb and Turner \cite{Kolb:1990vq}. All effects of the continuum physics are encoded in the calculation of $\langle \sigma v \rangle$.

\subsection{Freeze-Out in a Toy Model}

As an illustration, consider a toy model described by \cref{eq:action_scalar_continuum} and \cref{eq:Higgs_portal} where the DM and ``SM'' are both scalars, coupled through a 4-point coupling independent of the DM-mode masses. For simplicity, we set $m_h=0$ in this illustrative example. The tree-level matrix element in \cref{eq:sigma_v} is then simply ${\cal M}=\lambda$. An explicit calculation yields\footnote{We used an integral representation 
\bea
K_n (z) = \frac{\pi^{1/2} \left( \frac{z}{2} \right)^n}{\Gamma \left( n + \frac{1}{2} \right)} \int_1^\infty dt \; e^{-zt} \left( t^2 - 1 \right)^{n-1/2},
\eea
 Using this, it is straightforward to show that
\bea
\int d \Pi_\mu e^{- \beta E_\mu} = \frac{1}{4\pi^2} \mu \beta^{-1} K_1 (\mu \beta), \hspace{0.3cm} \int \frac{d^3 p}{(2\pi)^3} \; e^{-\beta E_\mu} = \frac{1}{2\pi^2} \mu^2 \beta^{-1} K_2 (\mu \beta).
\eea
The second identity is easily obtained by taking a partial derivative with respect to $\beta$ of the first identity and using $\frac{\partial \left( z^{-n} K_n (z) \right)}{\partial z} = - z^{-n} K_{n+1} (z)$. 
}
\beq
\langle \sigma v \rangle \,=\, \frac{\lambda^2}{32\pi}\,\left( \frac{I_1(\beta)}{I_2(\beta)}\right)^2.
\eeq{sv_tm}
Here we defined
\beq
I_n(\beta)\,\equiv\, \int \frac{d\mu^2}{2\pi} \rho(\mu^2)\,\mu^n\,K_n(\beta \mu)\,,
\eeq{Idef} 
where $K_n$ is the modified Bessel function of the second kind. In \cref{sv_tm} the $I_1^2$ in the numerator is the result of performing the phase space integrals for the continuum DM, while the $I_2^2$ in the denominator originates from the integrals corresponding to the $1/n_{\rm eq}^2$ factor in \cref{eq:sigma_v}.
As explained above, freeze-out occurs when the temperature drops below the gap scale, so we can approximate the Bessel functions at large argument. This yields 
\beq
I_n(\beta)\,\approx\, \sqrt{\frac{\pi}{2\beta}}\,\int_{\mu_0}^\infty \frac{d\mu}{\pi} \rho(\mu^2)\,\mu^{n+1/2}\,e^{-\beta\mu} \,.
\eeq{Iapprox}  
To make it even more explicit, let's assume a specific form for spectral density. Motivated by 5D model building (see \cref{sec:DM_spectral_density_5D}), we use 
\beq
\rho(\mu^2) =  \frac{\rho_0}{\mu_0^2} \,\left(\frac{\mu^2}{\mu_0^2}-1\right)^{1/2},
\eeq{rho0}
where $\rho_0$ is a dimensionless constant.
In \cref{sec:DM_spectral_density_5D} we show that this is indeed the form of the spectral density near the gap scale, and in \cref{appendix:spectral_density_near_gap} we present a more general argument for this.
With this assumption, the integrals in~\leqn{Iapprox} can be evaluated using the saddle-point approximation, giving\footnote{This result is valid as long as the saddle point is near the gap scale, $\mu_{\rm saddle} \approx \mu_0 + \mathcal{O} (T)$, and independent of $n$ of $I_n (\beta)$.}
\beq
\langle \sigma v \rangle \,=\, \frac{\lambda^2}{32\pi \mu_0^2} + \mathcal{O} \left( \frac{T}{\mu_0} \right) \,.
\eeq{sv_ttm}
Here, the term $\mathcal{O} (T/\mu_0)$ encodes the corrections due to the continuum nature of our DM and is typically expected to be of the order of $\sim 10 \%$. 
Note that the result is independent of $\rho_0$, and depends only weakly upon the assumed functional form of spectral density (although may need to be modified if it changes very rapidly or is very suppressed near the gap). 

We can also explicitly verify that assumptions made in the derivation of eq.~\leqn{n_evol} indeed hold in this toy model. Let $\Gamma_{\rm an}$ and $\Gamma_{\rm QES}$ denote the rates at which a DM state of energy $E$ undergoes annihilation and QES respectively. Estimating the rates in this model gives $\Gamma_{\rm an}\sim \Gamma_{\rm QES} \gg H$ for $T>\mu_0$ (assuming $\lambda\sim {\cal O}(1)$), so both reactions are active and maintain thermal and chemical equilibrium between SM and DM. However when $T<\mu_0$, the ratio $\Gamma_{\rm an} / \Gamma_{\rm QES}$, which is roughly the ratio of number density of non-relativistic state to that of relativistic state $n_{\rm NR} / n_{\rm R}$, becomes
\beq
\frac{\Gamma_{\rm an}}{\Gamma_{\rm QES}} \,\sim\,\left(\frac{\mu_0}{T}\right)^{3/2}\,e^{-\mu_0/T}\,, 
\eeq{Rratio}    
so that annihilations decouple well before QES, as assumed. At $T<\mu_0$, QES reaction with incoming DM mode of mass $\mu$ can have final-state DM state in the mass range $\left[ \mu_0 , \mu+T \right]$, so that chemical equilibrium among the DM modes of all possible masses is maintained. This justifies our assumption that the DM chemical potential $\eta$ is $\mu$-independent during the freeze-out.

\section{WIC model using the Vector Boson Portal}
\label{sec:Z_portal_model}

We are now ready to present a fully realistic WIC model based on the Z/W portal. In this section we will be discussing it based on a 4D description assuming that a gapped continuum mode is readily available and can be coupled to the SM. The resulting theory is presented in \cref{subsec:4D_toy_Z_portal}, while the evaluation of the dark matter relic density in this theory is in \cref{subsec:relic_density_Z-portal} below. A UV completion of this theory can be obtained using a warped extra dimensional construction with a non-trivial scalar field profile, which will be discussed in \cref{sec:DM_spectral_density_5D}.

\subsection{4D Effective $Z$-portal Model}
\label{subsec:4D_toy_Z_portal}

 We denote the field corresponding to the continuum DM by $\Phi$, which is assumed to be a complex scalar with no SM gauge quantum numbers.
 An exactly conserved $Z_2$ discrete symmetry is assumed, under which $\Phi$ is odd while all SM fields are even. This symmetry ensures DM stability. In this section, we describe $\Phi$ as a 4D field with an unusual ``kinetic term'' corresponding to the gapped continuum,  while  in the next section we will lift it to the boundary value of a 5D field. In order to  obtain non-vanishing interactions of the continuum DM with the SM $W,Z$ bosons we will mix it with another complex scalar field $\chi$ (with a canonical kinetic term)  which is a doublet under $SU(2)_L$, carries $U(1)_Y$ charge $-1/2$, and is odd under the $Z_2$.\footnote{One may wonder why we don't just couple $\Phi$ directly to $Z$ and $W$ by giving it SM gauge charges. In fact, this is possible. The mixing with mediator $\chi$ in this 4D construction is just to avoid continuum partners of $Z$ and $W$, which are required in the direct coupling case according to 5D consistency, and yield a more complicated theory.} This mixing is possible in the presence of the Higgs VEV. We assume that the $\chi$ field itself does not break the electroweak symmetry, hence its mass is a free parameter and can be taken to be relatively high $m_\chi \gg v$. The Lagrangian of the theory is 
\bea
&& {\cal L} = {\cal L}_{\rm SM} + {\cal L}_{\Phi} + {\cal L}_{\chi} + {\cal L}_{\rm int} \label{eq:Z-portal_4D} \label{eq:effective1} \\
&& {\cal L}_{\Phi} = \Phi^\dagger (p) \Sigma (p^2) \Phi (p) \label{eq:effective2}\\
&& {\cal L}_{\chi} = \left( D_\mu \chi \right)^\dagger \left( D^\mu \chi \right) - m_{\chi}^2 \chi^\dagger \chi \label{eq:effective3}\\
&& {\cal L}_{\rm int} \,=\, - \lambda \Phi \, \chi H\,+\, {\rm c.c.}   \label{eq:inter}
\eea
The quadratic action of $\Phi$ describes the continuum spectrum through the spectral density $\rho (p^2)$ defined by
\bea
\rho (p^2) = -2 {\rm Im} \Sigma^{-1} (p^2).
\eea
The covariant derivative appearing in ${\cal L}_{\chi}$ includes couplings to the SM $W$ and $U(1)_Y$ gauge bosons. We will assume that the temperature of the Universe is low enough that a Higgs VEV has already formed, $T < v$, and also $T<m_\chi$. 
When the Higgs gets a vev, ${\cal L}_{\rm int}$-term induces ``mass mixing'' between continuum states created by $\Phi$ and the neutral components of $\chi$.\footnote{Charged components of $\chi$ have mass $m_\chi\gg \mu_0$ and will not play a role in DM phenomenology.} In the standard case where $\Phi$ describes a single massive particle (as opposed to gapped continuum), the mass eigenstates would be given by the usual expressions
\beq
\tilde{\Phi} = \cos\alpha\,\Phi + \sin\alpha\,\chi^0,~~~~~\tilde{\chi}^0 = -\sin\alpha\,\Phi + \cos\alpha\,\chi^0. \nonumber
\eeq{rot}
The mixing angle would be
\beq
\tan 2\alpha \,=\,\frac{\sqrt{2} \lambda v}{m_\chi^2-m_\Phi^2}\,, \nonumber
\eeq{alpha}
where $v=246$ GeV. In the continuum case, we can get a similar result  by simply integrating out $\chi^0$ using its EoM (assuming $\chi$ is sufficiently heavy, in particular $m_\chi \gg \mu_0$ and the temperature is low enough that only states close to $\mu_0$ will be relevant).
In this case the EoM for $\chi^0$ implies 
\bea
\chi^0 = - \frac{\lambda v}{\sqrt{2}} \frac{\Phi}{\square + m_\chi^2}
\eea
where $\square \equiv \partial_\mu \partial^\mu$. 
Substituting this back into the action results in an effective action for $\Phi$ with mixing angles dependent on the mode mass $\mu$: 
\begin{equation}
\tan 2\alpha_\mu \,=\,\frac{\sqrt{2} \lambda v}{m_\chi^2-\mu^2}.
\label{eq:mu_dependent_mixing_angle}
\end{equation}
Since we are assuming that  $\mu \ll m_\chi$  for all relevant $\mu$ we may safely drop the $\mu$-dependence in the mixing angle.
The couplings of continuum modes with $p^2 = \mu^2$ to the SM $Z$ and $W$ gauge bosons are inherited from its mixing with $\chi^0$. 
The effective Lagrangian describing these couplings are given by (dropping the $\mu$-dependence in the mixing angle)
\begin{equation}
{\cal L}_{\Phi \text{-} Z,W} = \sin^2 \alpha \left[ -  \frac{i}{2} g_Z \left( \partial_\mu \Phi^\dagger \Phi - \Phi^\dagger \partial_\mu \Phi  \right)  Z^\mu + \frac{1}{4} g_Z^2 \Phi^\dagger \Phi Z_\mu Z^\mu + \frac{1}{2} g^2 \Phi^\dagger \Phi W^+_\mu W^{- \mu} \right]  
\label{eq:4D_Phi_ZW}
\end{equation}
where $g, g^\prime$ are the standard $SU(2)_L \times U(1)_Y$ couplings, and $g_Z = \sqrt{g^2+g'^2}$.
At first glance, this equation simply describes a complex scalar coupled to $Z$ and $W$ with extra factor of mixing angle $\sin^2 \alpha$. We emphasize, however, that $\Phi$ excites a whole set of free continuum states with the probability governed by the spectral density $\rho (\mu^2)$. Hence, \cref{eq:4D_Phi_ZW} contains couplings of the continuum modes for all values of $p^2 = \mu^2$  to $W$ and $Z$. 

In addition to the interaction term included in the Lagnarigian~\leqn{eq:inter}, a coupling $\lambda_\Phi |H|^2\Phi^2$ is also allowed by symmetries. In fact, even if not present in the original Lagrangian, this term will be induced at low energies by integrating out the $\chi$ field. This term induces a ``Higgs portal" interaction between the DM and SM. For $\mu_0<m_h/2$, it induces an exotic Higgs decay to two DM particles, which is constrained by the LHC data. This constraint is included in our analysis. We assume that the effects of the Higgs portal term in all other observables of interest are subdominant to those of the $Z$-portal interactions in~\leqn{eq:4D_Phi_ZW}. This assumption does not require strong tuning of $\lambda_\Phi$. For example, in relic density calculations, we need $\lambda_\Phi \lsim g \sin^2\alpha \sim 10^{-2}-0.1$ in the parameter region of interest. This constraint is further weakened for DM gap scale below the $W$ mass, thanks to extra Yukawa suppression in Higgs-mediated annihilation. An additional phenomenological constraint arises from the Higgs portal due to late-time DM decays DM$(\mu_1)\to$ DM$(\mu_2)+2\gamma$, mediated by the off-shell Higgs exchange. This decay (absent in the $Z$ portal model) can reionize hydrogen atoms after CMB decoupling, contrary to observations. This constraint rules out a model in which the Higgs portal is the dominant DM-SM interaction. However, if both $Z$ and Higgs portals are operational, the $2\gamma$ branching ratio is suppressed by an additional factor of $(\alpha/\pi)^2\sim 10^{-5}$ from the $h\gamma\gamma$ loop-induced vertex, which is sufficient to avoid this constraint. Beyond the Higgs portal, there may be additional non-renormalizable interactions between $\Phi$ and the SM. We assume that such interactions, if present, are generated at a scale well above electroweak, and thus their effects are negligible.

\subsection{Relic abundance of continuum $Z$-portal DM}
\label{subsec:relic_density_Z-portal}

In this section, we compute rates for processes relevant for the thermal freeze-out of continuum DM introduced above. 
We then show a region of parameter space $(\sin \alpha, \mu_0)$ that reproduces observed relic density. The spectral density is assumed to have the generic form in \cref{rho0}, although as remarked in \cref{sec:freeze-out_DM} the relic density is essentially independent of this assumption. 
 
Thermal-relic WIC occurs when the DM sector is in thermal and chemical equilibrium with SM at high temperature, at least $T \sim $TeV. This requirement places a lower bound on the effective coupling, and hence on $\sin \alpha$: 
\bea
\Gamma = \langle \sigma v \rangle n \sim \frac{g_Z^4 \sin^4 \alpha}{8\pi}  T \gtrsim \frac{T^2}{M_{\rm pl}} \;\;\; \to \;\;\; \sin \alpha \gtrsim 10^{-4} \left( \frac{{\rm TeV}}{M_{\rm pl}} \right)^{1/4}.
\eea
Below we assume that this bound is satisfied and thermal freeze-out occurs. The observed relic density indicates values of $\sin\alpha$ that are easily consistent with this bound, so the calculation is self-consistent.  

We now move on to the discussion of thermal freeze-out. Recall that the quasi-elastic scattering process which establishes thermal equilibrium between DM sector and SM decouples much later than the annihilations (see \cref{sec:freeze-out_DM}).
Therefore, when thermal freeze-out occurs we can safely assume that DM sector is in thermal equilibrium with SM. 

As discussed around \cref{n_evol} and \cref{eq:sigma_v} the Boltzmann equation for continuum is the same as particle DM except that, crucially, the thermal averaged rate includes averaging over the continuum spectrum.
The relevant annihilation processes depend on the size of gap scale $\mu_0$:
\begin{figure}
\begin{center}
\includegraphics[width=7.5cm]{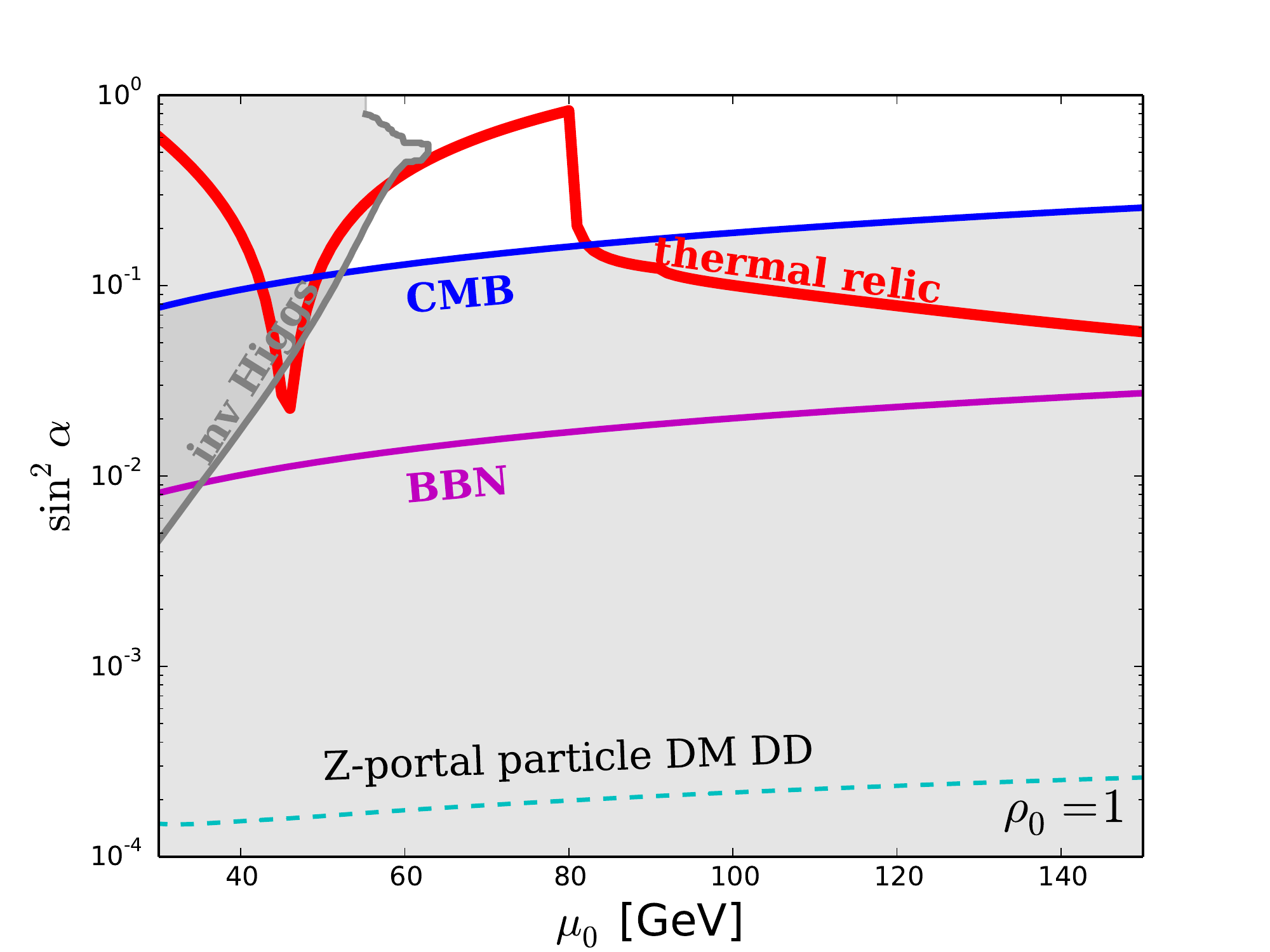} 
\includegraphics[width=7.5cm]{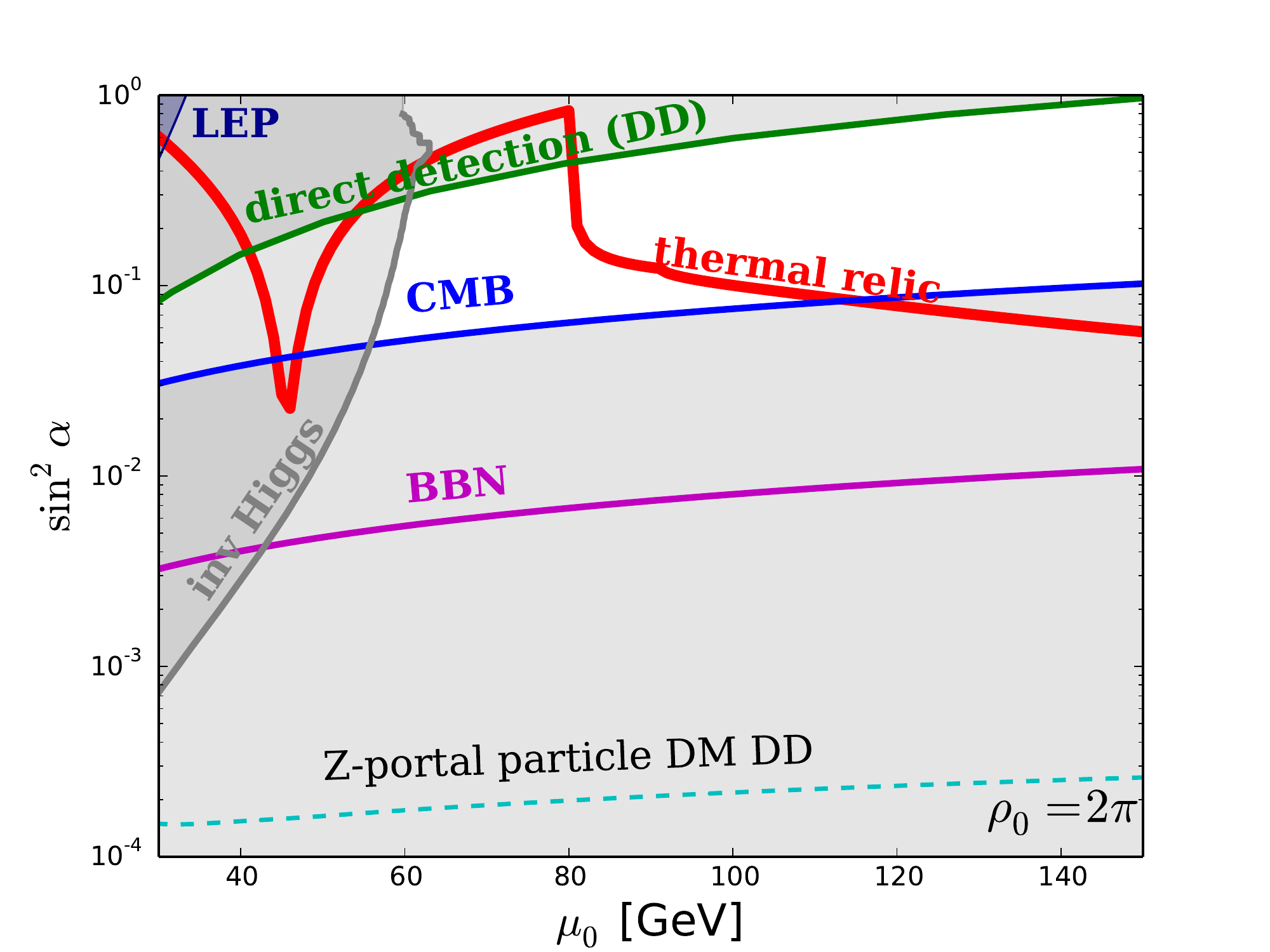} 
\end{center}
\caption{The parameter space of the Z-portal WIC DM, for $\rho_0=1$ (left) and $2\pi$ (right). 
The red curve corresponds to the observed relic density. 
The region below the blue (magenta) line is ruled out by the CMB (BBN) observations. 
The region above the green line is constrained by the XENON1T direct detection experiment \cite{Aprile:2018dbl}. For comparison, the direct detection constraint for Z-portal particle DM is shown in the cyan dashed line. The region above the gray line is ruled out by LHC constraint on exotic Higgs decays (for $m_\chi=500$~GeV). LEP bound from on-shell decays $Z\to\phi\phi^*$ is also shown. The details of the bounds are discussed in the companion paper~\cite{WIC_PRL}.}
\label{fig:Z-portal_relic} 
\end{figure}
\begin{itemize}
\item[(1)] $\mu_0 < m_W$ : The dominant process is $\phi \phi^* \to f \bar{f}$ via the s-channel $Z$ exchange. Here, $f$ denotes SM fermions (excluding top quark). The rate is given by 
\bea
\left\langle \sigma v \left( \phi \phi^* \to Z^{(*)} \to f \bar{f} \right) \right\rangle \approx \frac{g_Z^2 \sin^4 \alpha \; v_{\rm rel}^2}{128 \mu_0^2} \frac{\Gamma_Z}{m_Z} \left[ \left( 1 - \frac{m_Z^2}{4 \mu_0^2} \right)^2 + \frac{m_Z^2 \Gamma_Z^2}{16 \mu_0^4} \right]^{-1}.
\label{eq:sigma_v_ff}
\eea
The appearance of the relative velocity $v_{\rm rel}^2$ shows that this process is p-wave. The factor $\Gamma_Z / m_Z$ comes from the $Z \to f \bar{f}$ vertex of the Feynman diagram, and the last factor in the square bracket is the $Z$ propagator. When $\mu_0 \ll \frac{m_Z}{2}$, one sees that $\langle \sigma v \rangle \propto \frac{\sin^4 \alpha \mu_0^2}{m_Z^4}$, and so for the correct relic density $\sin \alpha$ decreases as $\mu_0$ increases. On the other hand, for $\frac{m_Z}{2} \ll \mu_0 < m_Z$, instead we get $\langle \sigma v \rangle \propto \frac{\sin^4 \alpha}{\mu_0^2}$. Hence, $\sin \alpha$ increases with $\mu_0$. These features are seen in \cref{fig:Z-portal_relic}. 
\item[(2)] $\mu_0 \sim m_W$ : In this regime, in addition to $\phi \phi^* \to f \bar{f}$, the three-body process, $\phi \phi^* \to W W^* \to W \ell \bar{\nu}$ can make a significant contribution. This, however, will be only relevant for $\mu_0$ very close to $m_W$ and we leave its explicit computation for a future investigation. For this reason, we warn that an $\mathcal{O} (1)$ (at most) correction may be required to the relic density curve in \cref{fig:Z-portal_relic} for $\mu_0 \approx m_W$. 
\item[(3)] $m_W < \mu_0 < m_Z$ : Now, a $W$ pair can be produced on-shell: $\phi \phi^* \to W^+ W^-$. The rate is estimated to be

\bea
&& \left\langle \sigma v \left( \phi \phi^* \to W^+ W^- \right) \right\rangle_{\rm ct} \approx \frac{g^4 \sin^4 \alpha}{128\pi \mu_0^2} \sqrt{1- \frac{1}{y}} \left( 4y^2 - 4y + 3 \right), \label{eq:sigma_v_WW_ct} \\ 
&& \left\langle \sigma v \left( \phi \phi^* \to W^+ W^- \right) \right\rangle_{\rm s} \approx \frac{g^4 \sin^4 \alpha \; v_{\rm rel}^2}{96 \pi \mu_0^2} \sqrt{1-\frac{1}{y}} \frac{4y^2+20y+3}{y^2} \left( 4 - \frac{1}{x} \right)^{-2}  \;\; \label{eq:sigma_v_WW_s}
\eea
where $y = \mu_0^2 / m_W^2$ and $x = \mu_0^2 / m_Z^2$. The first contribution comes from a contact interaction, while the second is from a s-channel $Z$ exchange.\footnote{The absence of the interference between the contact and s-channel contributions to the matrix element is due to the fact that the former is purely real while the latter is purely imaginary.}
Eq.~(\ref{eq:sigma_v_WW_ct}) is an s-wave process while Eq.~(\ref{eq:sigma_v_WW_s}) is p-wave. As $\mu_0 \gg m_W$, the rates become
$\langle \sigma v \rangle_{\rm ct} \propto \frac{\sin^4 \alpha}{m_W^4} \mu_0^2$ and $\langle \sigma v \rangle_{\rm s} \propto \frac{\sin^4 \alpha}{\mu_0^2}$. This means that at $\mu_0 \gg m_W^2$, the contact contribution dominates and $\sin \alpha$ drops with increased $\mu_0$ for fixed relic density. 
As $\mu_0$ is brought close to $m_W$ (so $y \approx 1$), both processes get phase-space suppression as captured by $\sqrt{1-1/y}$. The observed relic abundance then is achieved by taking larger and larger $\sin \alpha$. These features are all seen in \cref{fig:Z-portal_relic}.
\item[(4)] $\mu_0 > m_Z$ : Finally, for $\mu_0$ larger than the $Z$ mass, a pair of $Z$ bosons can be produced on-shell. This process is due to the contact interaction and the rate is  
\bea
\left\langle \sigma v \left( \phi \phi^* \to Z Z \right) \right\rangle \approx && \frac{g_z^4 \sin^4 \alpha}{256 \pi \mu_0^2} \sqrt{1-\frac{1}{x}} \left[ 4x^2 - 4x + 3 - 8x (x-1) \sin^2 \alpha \frac{{}}{{}} \right. \nonumber \\
&& \hspace{4cm} \left. + \frac{16x^2 (x-1)^2 \sin^4 \alpha}{(2x-1)^2}  \right].
\label{eq:sigma_v_ZZ}
\eea
Here, again, $x=\mu_0^2/m_Z^2$. This process is s-wave as $\langle \sigma v \rangle \propto v_{\rm rel}^0$. We also note that at large $x$, the expression in the square bracket behaves as $[ \cdots ] \to x^2 \cos^4 \alpha$. The growth with $x$ indicates that longitudinal $Z$ modes do not decouple. This is in fact required by the Goldstone boson equivalence theorem due to non-vanishing $\Phi\chi H$ coupling in this model. (If the DM were directly coupled under the gauge symmetry, we would have $\sin\alpha=1$ and in fact $\overline{|{\cal M}|^2}$ does not grow at large $x$ in this limit, as expected.) For $x \gg 1$, the rate becomes $\langle \sigma v \rangle \propto \frac{\sin^4 2\alpha}{m_Z^4} \mu_0^2$, same scaling as in the case of $WW$ final state above. 
\end{itemize}

The region of parameter space where the observed relic density is reproduced is shown in \cref{fig:Z-portal_relic}: roughly, for $\mu_0\sim 100$ GeV, the effective coupling $g_{\rm eff} \approx g \sin^2 \alpha \gtrsim 10^{-2}$ is required. This is similar to ordinary particle DM with a Z-portal. The crucial difference is that while for ordinary particle the coupling of this size is completely ruled out by direct detection experiments, in the case of WIC this bound is much weaker due to the suppression of direct detection rates discussed in \cref{sec:preview}. The bounds from direct detection, as well as from the consistency of ionization history of the universe in the presence of late-time WIC decays, are discussed in the companion paper~\cite{WIC_PRL} and are summarized in \cref{fig:Z-portal_relic}. Z-portal WIC dark matter is consistent with all experimental bounds.

\section{Continuum Spectral Density and UV-Complete WIC Model from a Warped Spacetime}
\label{sec:DM_spectral_density_5D}

Finally we present a realistic implementation of the continuum DM scenario with a Z-portal in a local and unitary 5D theory. The main goal is to construct a UV completion of the 4D effective theory shown in \cref{subsec:4D_toy_Z_portal} and used for the relic density calculation in \cref{subsec:relic_density_Z-portal}. We start by recalling the soft-wall construction of a gapped continuum in 5D warped space following~\cite{Cabrer:2009we}. In \cref{subsec:5Dspectral} we examine the detailed properties of the resulting spectral density, paying particular attention to the behavior of $\rho$ close to the gap, which turns out to be crucial for the dark matter phenomenology. Finally in \cref{subsec:5D_Z_portal}  we present an explicit 5D model whose 4D effective theory (by integrating out the bulk) matches to the effective theory given in \cref{subsec:4D_toy_Z_portal}, and which  makes a concrete prediction  for the form of the spectral density $\rho (\mu^2)$. 
The  holographic dual description of our 5D gapped continuum in terms of strongly coupled CFT is discussed in \cref{app:AdS_CFT}.

\subsection{The Cabrer-von Gersdorff-Quiros (CGQ) Background}
\label{subsec:CGQ}

In the warped 5D setup we will have a 3-brane placed at the position $z=R$, which from the point of view of the bulk field will be a UV brane cutting off the space. Toward the IR the extra dimension is  non-compact, supplemented by a \emph{background scalar field} $\varphi (y)$, whose back-reaction is responsible for a so-called soft-wall, resulting in a finite proper length for the extra dimension. Note that in this paper we are not trying to solve the Higgs hierarchy problem, but merely provide a complete construction of a gapped continuum. Hence our UV scale defined by the location of the UV brane is not exponentially larger than the weak scale, but rather comparable to it. (If we did try to embed this setup into a traditional Randall-Sundrum-type warped extra dimensional model \cite{Randall:1999ee} we would need to extend it beyond our UV brane. In fact in that UV complete theory the brane we are using as a UV cutoff here would actually be identified with the traditional IR brane of RS, and a new UV brane would have to be introduced in the far UV at scales exponentially higher than the weak scale.) Another point we want to mention is that below we will work in zero temperature background. In the context of cosmology, such a zero temperature geometry will arise after a thermal phase transition around $T \lesssim T_c$ (assuming not highly super-cooled phase transition). In dual CFT description, at $T > T_c$, the theory is in hot CFT phase (dual to AdS black hole phase in 5D), while at $T < T_c$ it is in gapped continuum phase. A non-trivial cosmology of gapped continuum is possible provided $\mu_0 < T_c$ which we assume.  
We note that at finite $T < T_c$, the geometry is not quite yet that of zero temperature, but rather thermal AdS where the temporal direction is compactified with radius $\sim 1/T$. It is then expected that spectral density computed in such a background will exhibit $T$-dependence. On the other hand, naive dimensional analysis suggests that thermal corrections are suppressed as $\mathcal{O} (T/T_c)$. In this work, we ignore such thermal corrections.

The 5D action of the coupled scalar-gravity system is given by
\begin{equation}
S=\int d^5 x\sqrt{g} \left( -M^3 R +\frac{1}{2} g^{MN} (\partial_M \varphi) (\partial_N \varphi) - V(\varphi ) \right)-\int d^4x  \sqrt{g^{\text{ind}}} \; V_4 (\varphi )
\end{equation}
where $M^3$ is the 5D Planck mass and we are using the metric signature $(+,-,-,-,-)$. 
\footnote{Since our spacetime manifold has a boundary, the Gibbons-Hawking-York (GHY) boundary term is required so that the variational principle for gravity is well-defined~\cite{York:1972sj, Gibbons:1976ue}. Its explicit form is not needed for our discussion and we ignore it. }

Using the proper distance as the coordinate along the extra dimension we parametrize the metric as 
\begin{equation}
\diff s^2=e^{-2A(y)}\diff x^2-\diff y^2
\end{equation}
where $A(y)$ is the warp factor. While solving the coupled Einstein-scalar equations analytically in general is rather challenging, there is a special case when the coupled second order equations simplify to first order ordinary differential equations which can be analytically solved. Such a simplification occurs when the scalar potential can be given in terms of a superpotential $W$ via the relation~\cite{DeWolfe:1999cp,Csaki:2000wz} 
\bea
V (\varphi) = \frac{1}{8} \left( \frac{\partial W}{\partial \varphi} \right)^2 - \frac{1}{12 M^3} W^2.
\eea
In terms of the superpotential, the bulk EoMs take a simple form
\bea
\frac{d \varphi}{d y} = \frac{1}{2} \frac{\partial W}{\partial \varphi}, \;\;\;\; \frac{d A}{d y} = \frac{1}{12 M^3} W.
\eea
The superpotential leading to the desired 5D background\footnote{Note that this is a special case of a class of superpotentials parametrized as $W= 12k M^3 \left( 1+ e^{\nu \varphi / \sqrt{6M^3}} \right)$. For $\nu > 1$ one has a discrete spectrum, while for $\nu<1$ a continuum without a gap. The critical value $\nu =1$ corresponds to a gapped continuum of the sort we are considering.} is given by~\cite{Cabrer:2009we}, 
\bea
W (\varphi) = 12 k M^3 \left( 1 + e^{\varphi/ \sqrt{6M^3}} \right)
\eea
where $k$ is the AdS curvature scale asymptotically away from the soft wall. The solution of the first order differential equations yield the background 
\begin{equation}
A(y)=ky - \log{\left (1-\frac{y}{y_s}\right )},
\label{eq:A(y)}
\end{equation}
and
\begin{equation}
\varphi (y)= - \sqrt{6M^3} \log{\left ( k (y_s - y) \right )}.
\label{eq:phi(y)}
\end{equation}
It is observed that there is a singularity located at $y_s$ and it corresponds to the finite distance location of the curvature singularity where the spacetime ends in the $y$ coordinates (corresponding to $z\to \infty$ in the conformally flat coordinates). It is also seen that for $y \ll y_s$, $A \to ky$ and the geometry is just $AdS_5$. The beauty of this solution is that it fully includes the backreaction of the metric to the presence of the scalar field - which is indeed the origin of the actual curvature singularity. 

\subsection{Realistic Gapped Continuum Spectral Density}
\label{subsec:5Dspectral}

The gapped continuum is obtained by considering additional fields (scalar, vector or fermion) in this background\footnote{For the case of the fermions, one needs to introduce an additional Yukawa-like coupling to the background scalar field $\varphi$ to yield a gapped continuum. Without such coupling the continuum would start at zero.}.
To be concrete, we consider the simplest case of  a scalar continuum with a mass gap, by introducing an additional scalar field $\Phi$ (which would play the role of dark matter) in this 5D set-up with the assumption of a stabilizing symmetry. We take this symmetry to be a discrete $Z_2$, under which $\Phi$ is odd. The Lagrangian for this additional scalar is
\begin{equation}
\mathcal{L}=\sqrt{g}\left[g^{MN}D_M\Phi^\dagger D_N\Phi-m^2|\Phi|^2\right ],
\label{eq:bulk_Phi_action}
\end{equation}
where for simplicity we choose to set brane localized potentials to zero and ignore scalar self-interaction terms in the bulk. By means of integration by parts along the fifth dimension, the bulk action can be written so that it is proportional to the bulk EoM. This is a useful representation since it leads to a vanishing bulk action once the bulk field is evaluated at its classical solution. The integration by parts, however, induces a UV-localized term
\begin{equation}
\Delta S_{\rm UV} = \int_{\rm UV} d^4x \; e^{-4A} \Phi^\dagger \partial_y \Phi,
\label{eq:Delta_S_UV}
\end{equation}
which then turns into the (holographic) effective action~\cite{Barbieri:2003pr,Agashe:2007mc} once the bulk is integrated out at tree level.

Using a field redefinition $\Psi (p, y)=e^{-2A(y)}\Phi (p, y)$ where $p = \sqrt{p^2}$, the bulk EoM in the mixed momentum-position coordinates becomes
\begin{equation}
\left(-\partial_y^2+\hat{V}(y)\right)\Psi (p, y)=e^{2A(y)} p^2 \Psi (p, y)
\label{eq:ycoord}
\end{equation}
where the potential $\hat{V}(y)$ is given in terms of the warp factor by
\begin{equation}
\hat{V}(y) = m^2 + 4\left (A'(y)\right )^2 - 2A''(y).
\end{equation}
Here $()'$ denotes the derivative with respect to $y$.
Further insight into the modes in this potential can be gained by transforming \cref{eq:ycoord} into a 
Schr{\"o}dinger form, which can be achieved by going into the conformally flat $z$ coordinates
via $dz/dy = e^A$ and an additional rescaling $\psi = e^{A/2} \Psi$. 
In this frame, the bulk EoM turns into the standard Schr{\"o}dinger equation
\begin{equation}
- \ddot{\psi} + V(z) \psi = p^2 \psi
\label{eq:bulk_EoM_in_z}
\end{equation}
with the potential given by
\begin{equation}
V(z) = m^2 e^{-2A} + \frac{9}{4} \left( \dot{A} \right)^2 - \frac{3}{2} \ddot{A}.
\label{eq:V(z)}
\end{equation}
Here $\dot{()}$ denotes the derivative with respect to the conformal coordinate $z$.
An explicit expression for $V$ can be obtained using \cref{eq:A(y)}
\begin{equation}
V(z) = \frac{e^{-2ky}}{4 y_s^2} \left[ 4 m^2 (y_s - y)^2  + 15 \left( 1 + k (y_s - y) \right)^2 - 6\right].
\label{eq:V(z)_explicit}
\end{equation}
It is understood that $y$ should be expressed in terms of the conformally flat coordinate $y(z)$ via the transformation $dz/dy = e^A$. The potential approaches a constant
\begin{equation}
\mu_0^2 = \frac{9}{4 y_s^2}e^{-2 k y_s}
\label{eq:mu_0_5D}
\end{equation}
for $y\to y_s$, providing the mass gap for the continuum. 
Once the solution to the EoM, \cref{eq:bulk_EoM_in_z} and \cref{eq:V(z)}, is found, we obtain the boundary (or holographic) effective action. The 5d field $\Phi (x,z)$ can be written in terms of the source field in momentum space as $\Phi(p,z)= f(p,z) \hat{\Phi}(p)$ where $f(p,z)$ is the wave function related to the functions $\psi$ satisfying the simple Schr{\"o}dinger-type equation as $f(p,z)=e^{3A/2} \psi (p,z)$. Using this definition the holographic effective action is
\begin{align}
S_{\rm eff} &= \int_{\rm UV} d^4 x \; e^{-4A} \Phi^\dagger (x, y) \partial_y \Phi (x, y) |_{y=0} \nonumber \\
&= \int_{\rm UV} \frac{d^4 p}{(2\pi)^4} \;  \hat{\Phi}^\dagger (p)  \left( e^{-3A(z)}  \frac{f'(z,p)}{f(R,p)} \right)_{z=R} \hat{\Phi} (p)
\label{eq:holo_effective_action_derivation}
\end{align}
%
%
%
which is the final 5D expression for the effective action. We would like to translate this to our 4D effective theory in eqs.~(\ref{eq:effective1})-(\ref{eq:inter}). For this we need to also write the proper 5D version of the localized SM terms 
\begin{equation} 
\int d^4x \sqrt{g} (D_\mu H D^\mu H e^{2A} + D_\mu \chi D^\mu \chi e^{2A} -\hat{\lambda} k^\frac{1}{2} \Phi \chi H +h.c.)_{z=R}
\end{equation}
where $\hat{\lambda}$ is a dimensionless number, and we have used the AdS curvature scale $k$ to make up for the dimension of the coupling. In order to get the proper 4D effective action with an effective $\lambda = \hat{\lambda} k e^{-A}$ of the order of the electroweak scale we need the field redefinitions $H\to H e^{-A}, \chi \to \chi e^{-A}, \Phi \to \Phi e^{-\frac{3}{2} A} \sqrt{R}$. This will result in the 5D prediction of the effective 4D kinetic function \cref{eq:action_scalar_continuum} from  \cref{sec:physics_gapped_continuum} to be
\begin{equation}
\Sigma (p) = \frac{1}{R} \left.  \frac{f'(z,p)}{f(R,p)}\right\vert_{z=R} ,
\label{eq:Sigma(p)_5D}
\end{equation}
where $R$ is the location of the brane $R^{-1}= k e^{-A}$ for AdS-like metrics with $e^{A} = z k$.  
  Note that the units of the kinetic function are set by $\Sigma \sim 1/R^2$ the location of the brane. 
We emphasize again that this effective action was obtained starting from a local and unitary scalar field theory propagating in a self-consistent 5D background space, and should automatically be yielding a consistent effective 4D theory.

We can now find the expression for the spectral density close to the mass gap $\mu_0$. 
While the above potential $V (z)$ cannot be obtained analytically, the asymptotic form of the potential for $z\rightarrow \infty$ needed for the spectral density near the mass gap can be found explicitly.
As $y \to y_s$, we may get the expression for $k (y_s - y)$ in terms of $z$ by noting that the integrand in
\bea
z = \int dy \frac{y_s}{y_s - y} e^{ky}
\eea
has a rapidly varying factor $1/(y_s-y)$ and we may treat $e^{ky}$ to be approximately constant. Performing the integral with this approximation yields
\bea
z = - y_s e^{ky_s} {\rm log} (k (y_s-y)) = - \frac{3}{2 \mu_0} {\rm log} (k (y_s-y)) \;\; \to \;\; k (y_s-y) = e^{- \frac{2}{3} \mu_0 z}
\eea
where we used \cref{eq:mu_0_5D}.
Hence, the potential \cref{eq:V(z)_explicit} can be written near $z \to \infty$ as
\begin{equation}
V(z) \to \mu_0^2 \left[ 1 + \frac{10}{3} e^{-\frac{2}{3} \mu_0 z} + \left( \frac{4m^2}{9 k^2} + \frac{15}{9} \right)  e^{-2 \frac{2}{3} \mu_0 z} \right].
\label{eq:V(z)_z_infty}
\end{equation}
The solution of the Schr\"{o}dinger equation for the asymptotic potential (assuming an outgoing wave boundary condition $\psi \to e^{ikz}$ for $z\to \infty$, and setting the bulk mass to zero for simplicity) 
can also be explicitly found in terms of a generalized Laguerre polynomial: 
\bea
\psi(z,\mu) = D \,L_l^n ( 3\sqrt{5}e^{- 2 z \mu_0/3}) \exp \left(\frac{3}{2} \sqrt{1-\frac{\mu^2}{\mu_0^2}} \log \left(e^{-\frac{2
   \mu_0 z}{3}}\right)-\frac{3\sqrt{5}}{2} e^{-\frac{2 \mu_0 z}{3}}\right),
\eea
with $l=-(3\sqrt{5}+1)/2-3/2\sqrt{1-\mu^2/\mu_0^2}$, $n=3\sqrt{1-\mu^2/\mu_0^2}$, and an arbitrary coefficient $D$ is fixed by the normalization condition.
The spectral density is  related to the kinetic function $\Sigma (p)$ given in \cref{eq:Sigma(p)_5D}) as  
\begin{align}
\rho(p) = - 2 {\rm Im}\, \Sigma(p)^{-1}.
\end{align}
Since the potential in \cref{eq:V(z)_z_infty} is only valid for large $z \to \infty$ which is relevant for modes $\mu^2 \sim \mu_0^2$ we can expand the arguments of the Laguerre polynomial around the mass gap (with an expansion $\sqrt{\mu^2/\mu_0^2-1} \ll 1$) to obtain the approximate form of the spectral density near the mass gap $\mu \approx \mu_0$, 
\begin{equation}
\rho(\mu^2) =  \frac{\rho_0}{\mu_0^2}\left(\frac{\mu^2}{\mu_0^2}-1\right)^{1/2}\ ,
\label{eq:rho_near_gap}
\end{equation}
where $\rho_0$ is a dimensionless constant.
We can in fact show in a model independent way that around the mass gap this is indeed the expected form of the spectral density in a very general case. The one assumption we have to make is that for large $z$ the potential is well-approximated by a constant (which was clearly the case in the concrete 5D model investigated above). This discussion is presented in \cref{appendix:spectral_density_near_gap}.

\begin{figure}[ht]
\begin{center}
\includegraphics[width=10cm]{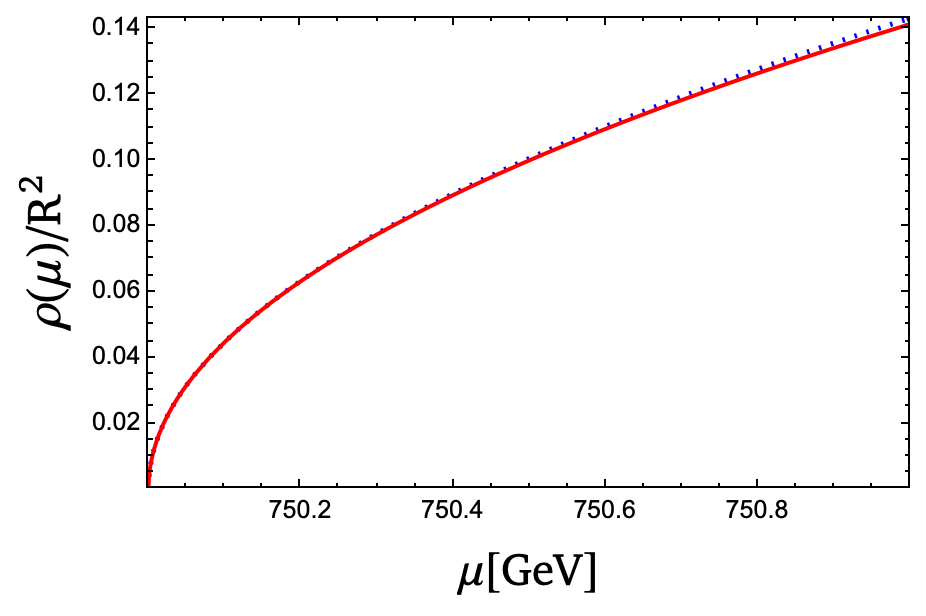} 
\end{center}
\caption{The shape of the spectral density near the gap scale $\mu_0$. For this plot, we choose $\mu_0 = 750$ GeV, $m=0$, and $R^{-1}=300$ GeV. The red solid curve is from the exact numerical solution and the blue dashed curve is a fit by a function $\rho (\mu^2) = \rho_0 / \mu_0^2 \left( \mu^2 /\mu_0^2-1\right)^{1/2}$ with a dimensionless normalization parameter $\rho_0$. We have fixed $k=10^{18}$ GeV and assumed a vanishing bulk scalar mass $m=0$. }
\label{fig:SD1} 
\end{figure}

\begin{figure}
\begin{center}
\includegraphics[width=10cm]{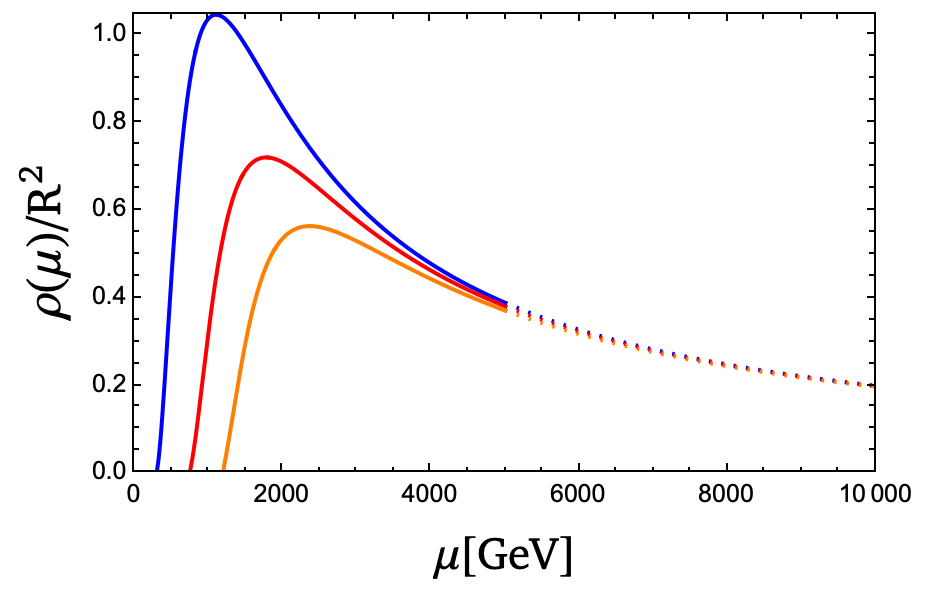} 
\end{center}
\caption{In this plot, we show the numerically obtained spectral density function over the full energy range assuming mass gaps: 
$\mu_0=$300 (Blue), 750 (Red), and 1200 (Orange) GeV, again for $k=10^{18}$ GeV and $m=0$. Below the cutoff of order $\mathcal{O}$ (TeV), we show $\rho (\mu)$ in solid curves, while above the cutoff they are shown as dashed curves. As usual, $\rho$ above the cutoff is not supposed to be used in effective theory calculations. Nevertheless, it is instructive to observe that $\rho$ exhibits universal behavior, i.e.~that of CFT.
}
\label{fig:SD2} 
\end{figure}

While the wave function and spectral density for general $\mu^2$ cannot be analytically determined,
we can nonetheless solve the Schr\"{o}dinger equation numerically (with the outgoing wave boundary conditions). It can be done either in the $z$ or $y$ coordinates (for the latter case the outgoing BC should be imposed very close to the singularity $y\to y_s$). This confirms the expression of the spectral density around the gap in \cref{eq:rho_near_gap}, which is illustrated in fig.~\ref{fig:SD1}. In fig.~\ref{fig:SD2} we show the entire spectral density function obtained from numerically solving the Schr\"{o}dinger equation.

\begin{figure}
\begin{center}
\includegraphics[width=10cm]{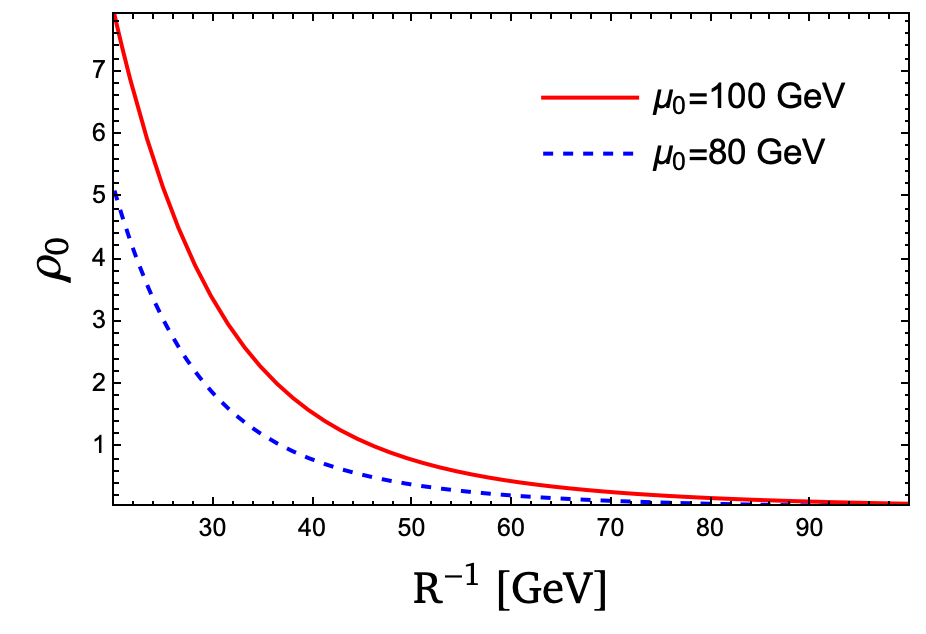} 
\end{center}
\caption{In this plot, we show the numerically obtained spectral density normalization $\rho_0$ as a function of the location of the brane $R^{-1}$. We have again fixed $k=10^{18}$ GeV and $m=0$.}
\label{fig:rho0} 
\end{figure}

We close this discussion by describing the overall normalization $\rho_0$ of the spectral density. As explained above this is fixed by identifying the 4D action (\ref{eq:action_scalar_continuum}) with the holographic effective action, using normalization that will produce our 4D effective theory from eqs.~(\ref{eq:effective1})-(\ref{eq:inter}).   
 The resulting $\rho_0$ will depend on the details of the model: as we have seen $\Sigma \sim 1/R^2$, hence we expect $\rho_0 \sim (\mu_0 R)^2$. In order to obtain $\rho_0 \sim {\cal O} (1)$ one needs $R^{-1} \sim \mu_0$, which is numerically verified in fig.~\ref{fig:rho0}, where we plot $\rho_0$ as a function of the position of the brane $R$. The requirement of $\rho_0 \sim {\cal O}(1)$ which is needed to obtain phenomenologically interesting models will imply a tuning of order $(\mu_0/{\rm TeV})^2$, of the percent level for this simplest model with a single bulk scalar playing the role of the gapped continuum.

\subsection{5D $Z$-portal Model}
\label{subsec:5D_Z_portal}

We are now ready to present the full 5D construction that incorporates the interactions of the continuum DM with the $W$ and $Z$. 
We start with the same setup as previously in \cref{subsec:CGQ} and \cref{subsec:5Dspectral} and assume the background as in \cref{eq:A(y)} and \cref{eq:phi(y)}. The SM is localized on the UV brane. The action for the scalar dark matter $\Phi$ reads
\bea
&& S = S_{\rm bulk} + S_{\rm UV} \\
&& S_{\rm bulk} = \int d^4 x dy \; \sqrt{g} \left( g^{MN} \left( \partial_M \Phi \right)^\dagger \left( \partial_N \Phi \right) - m^2 \vert \Phi \vert ^2 \right) \\
&& S_{\rm UV} = \int_{\rm UV} d^4 x \; \sqrt{g}  \left( {\cal L}_{\rm SM} + g^{MN} D_M \chi^\dagger D_N \chi  - \hat{m}_{\chi}^2 \vert \chi \vert^2 - \hat{\lambda} k^\frac{1}{2} \Phi \chi H + {\rm h.c.} \right).
\eea
As explained below \cref{eq:bulk_Phi_action} in \cref{subsec:5Dspectral}, using integration by parts, the bulk action can be rewritten so that the integrand is proportional to EoM of $\Phi$, but with an extra boundary term induced on the UV brane given in \cref{eq:Delta_S_UV}.
We have also seen that the bulk EoM takes the form of a standard Schr{\"o}dinger equation in terms of a new variable $\psi = e^{-\frac{3}{2} A} \Phi$ and conformal coordinate $z$ related to $y$ via $dz/dy = e^A$. The profile of $\Phi$ is given by $f=\psi e^{\frac{3}{2}A}$.
The potential \cref{eq:V(z)_explicit} approaches a constant value at large $z$, see \cref{eq:mu_0_5D}, revealing that the spectrum consists of a continuum starting at the gap scale $\mu_0$.
Once the solution for the ``profile'' $f(z,p)$ is found, we can integrate out the bulk by substituting $f$ back into the action. The bulk action vanishes trivially since it is directly proportional to the EoM. After proper rescaling as explained in the previous section we obtain the boundary (or holographic) effective action (see also \cref{eq:holo_effective_action_derivation}). 
\bea
S_{\rm eff} & = & \int \frac{d^4 p}{(2\pi)^4} \; \frac{1}{R}  \hat{\Phi}^\dagger (p) \left. \frac{f'(z,p)}{f(R,p)} \right\vert_{z=R} \hat{\Phi} (p) \nonumber \\ 
&& + \int d^4 x \;  \left( {\cal L}_{\rm SM} + \vert D_\mu \chi \vert^2 - m_{\chi}^2 \vert \chi \vert^2 - \hat{\lambda} R^{-1} \hat{\Phi} \chi H + {\rm h.c.} \right)
\eea
where the quadratic action of $\hat{\Phi}$ is expressed in momentum space since it is non-analytic in general.
As promised  this effective action is reproducing the 4D effective model eq.~(\ref{eq:effective1})-(\ref{eq:inter}) with the identifications 
\begin{align}
\Sigma (p^2) &= \frac{1}{R}  \left. \frac{f'(z,p)}{f(R,p)} \right\vert_{z=R} \label{eq:Sigma_from_5D} \\
\lambda &= R^{-1} \hat{\lambda}. \nonumber
\end{align}

Here, we started with a local and unitary microscopic theory in 5D, which predicts a specific form of $\Sigma (p^2)$ as given in \cref{eq:Sigma_from_5D}.
While the explicit form of the spectral density for arbitrary $p^2$ is not easy to work out analytically, nonetheless it is straightforward to find it numerically. Moreover, importantly, the form of the spectral density near the gap scale $\mu_0$ takes a universal form, \cref{eq:rho_near_gap}, as shown in \cref{subsec:5Dspectral} through an explicit 5D calculation. We also provide a more general argument in \cref{appendix:spectral_density_near_gap}.

The couplings of the DM modes with the SM weak and hyper-charge gauge bosons needed to study phenomenology can be obtained as described in \cref{sec:Z_portal_model}. The final results are simply \cref{eq:mu_dependent_mixing_angle} and \cref{eq:4D_Phi_ZW}. The dependence on $\rho (\mu^2)$ comes in when we compute rates using the formalism presented in \cref{sec:physics_gapped_continuum}.

\section{Conclusion and Outlook}
\label{sec:conclusion}

We presented a novel type of DM model, where the role of the dark sector is played by a Weakly Interacting Continuum (WIC). The continuum is assumed to be gapped at the weak scale, and interact with the SM EW sector, producing a continuum version of standard WIMP models. The continuum kinematics ensures that direct detection processes are strongly suppressed compared to familiar WIMPs, while in many other respects (relic abundance, indirect detection and some of the collider bounds) WIC DM is very similar to WIMPs. The suppression of the direct detection bounds re-opens the possibility of viable Z-portal DM models\footnote{For recent work on particle Z-portal DM with CP violation in the dark sector, see Ref.~\cite{Keus:2019szx}}. An unusual distinguishing aspect of WIC DM is the appearance of late decays of the sort ${\rm DM} (\mu_1)\to {\rm DM} (\mu_2)+ {\rm SM}$. Bounds on these late decays provide a lower bound on the interaction strength, leading to a well-defined allowed band in the parameter space (see \cref{fig:Z-portal_relic}). 
In addition, the WIC model has spectacular collider signatures driven by cascade decays of the continuum DM states produced at a collider.  

In this paper we have focused on providing the details of the underlying construction both of the continuum itself as well as its interactions with the SM. We have carefully defined the structure of generic free continuum field theories, which are subsequently coupled to the EW sector of the SM. We presented the structure of the Hilbert space for the free continuum and the basic elements of thermodynamics involving such states, which allowed us to derive the appropriate Boltzmann equation. We showed how to couple the continuum to the EW sector of the SM via a Higgs induced mixing. Using a simple effective theory of the continuum, we were able to calculate the relic density from the freeze-out of the WIC. A complete realistic model was obtained by considering a scalar field in a soft wall background in warped extra dimensions. This allowed us to find a concrete expression for the spectral density of a gapped continuum in a fully self-consistent theory, and verify the general form of the spectral density around the gap. The actual coupling to the SM is induced on the UV brane, providing a full implementation of the effective theory examined earlier. The full analysis of the phenomenology of this model will be presented in the companion paper~\cite{WIC_PRL}.

In summary, we showed that a dark sector described by a gapped continuum QFT can provide a fully realistic dark matter candidate, with  unique phenomenological features qualitatively different from any particle DM model. This opens up a new direction in DM model-building. While here we focused on weak-scale DM with a Z-portal, the idea can be applied to many other contexts, such as axionic DM, light thermal relics at the keV-GeV scales such as SIMPs, and so on. We look forward to further exploration in this direction.

\section*{Acknowledgments}
We are grateful to Barry McCoy, Eun-Gook Moon, Michele Redi, Carlos Wagner, Liantao Wang and Kathryn Zurek for helpful discussions.
C.C., S.H. G.K. and M.P. were supported in part by the NSF grant PHY-2014071. C.C. was also supported in part  by  the US-Israeli BSF grant 2016153.
S.H.\  was also supported by the DOE grants DE-SC-0013642 and  DE-AC02-06CH11357 as well as a Hans Bethe Post-doctoral fellowship at Cornell.
G.K. is supported by the Science and Technology Facilities Council with Grant No. ST/T000864/1.
S.L.\ was supported by the Samsung Science and Technology Foundation.
W.X. was supported in part by the DOE grant DE-SC0010296.

\appendix

\section{Gapped Continuum from Five-Dimensional Flat Space}
\label{app:flat_X_dim}
	
In this appendix we consider an infinite, flat 5D space with coordinates $(x^\mu, z)$. A 5D scalar field $\Phi(x^\mu, z)$ with mass $m_0$ propagates on this space. We discuss how this theory can be alternatively described as a 4D theory with a gapped continuum spectrum. We should note that this setup cannot be used to construct realistic models of the kind we consider in the paper, because gravity remains five-dimensional at all distance scales. Still, it is a useful example to consider to gain intuition about gapped continuum spectrum and spectral density in a simple context.

\subsection{Infinite 5D and Spectral Density}
	
The scalar propagator has the form
	\beq
	\langle \Phi(x^\mu, z) \Phi(0)\rangle \,=\,\int \frac{d^5 P}{(2\pi)^5}\,\frac{i}{P^2-m_0^2+i\epsilon}\,e^{-i(p\cdot x - z k) }\,,  
	\eeq{5Dprop}  
where $P=(p^\mu, k)$. Fourier transforming into momentum space along the four $x^\mu$ dimensions, we get
\beq
\Pi(p^2, z) \,=\, \int \frac{dk}{2\pi}\,\frac{i}{p^2-(k^2+m_0^2)+i\epsilon}\,e^{i z k }\,.
\eeq{Mixprop}	
Consider a 4D ``brane'' at $z=0$. The brane-to-brane propagator is 
\beq
\Pi(p^2, 0) \,=\, \int \frac{dk}{2\pi}\,\frac{i}{p^2-(k^2+m_0^2)+i\epsilon}\,.
\eeq{Mixprop0}	
Defining $s=k^2+m_0^2$, we can rewrite 
\beq
\Pi(p^2, 0) \,=\, \int_{m_0^2}^{+\infty} \frac{ds}{2\pi}\,\frac{i}{p^2-s+i\epsilon}\,\rho(s) \,,
\eeq{SD}	
where 
\beq
\rho(s) = \frac{1}{2\sqrt{s-m_0^2}}
\eeq{rho}  	
is the spectral density. Thus we have recast this trivial 5D theory as a 4D theory with gapped continuum spectrum and a non-trivial spectral density.  

\subsection{Compactified Theory and KK Picture}

Let us now consider the same theory with the $z$ direction compactified on a circle of radius $R$. This gives a familiar KK theory.
The scalar field decomposes as $\Phi(x, z) = \sum_n f_n(z) \phi_n(x)$, where $\phi_n(x)$ are 4D fields with masses
\beq
m_n^2=m_0^2 + \left(\frac{n}{R}\right)^2.
\eeq{KKmodes}	 
and $f_n=\frac{1}{\sqrt{2\pi R}}\,\cos (n\pi z/R)$. The brane-to-brane propagator can then be expressed as 
\beq
\langle \Phi(x^\mu, 0) \Phi(0)\rangle \,=\,\sum_{m,n} f_n(0)f_m(0) \, \langle \phi_n(x) \phi_m(0) \rangle\,,
\eeq{PropKK}
which gives
	\beq
	\Pi(p^2, 0) = \frac{1}{2\pi R} \sum_n \frac{i}{p^2-m_n^2+i\epsilon}.
	\eeq{PropKK1}
	Now consider going back to the infinite 5D theory by taking the limit $R\to\infty$. We expect that in this limit the sum over $n$ turns into an integral over KK mass, labeled by a continuum parameter $\mu^2$. For large $R$, the splitting between neighboring KK modes is
	\beq
	\Delta \mu^2_n \equiv m_{n+1}^2-m_n^2 \approx \frac{2n}{R^2} = \frac{2}{R}\sqrt{\mu^2-m_0^2}.
	\eeq{split}
	Thus,
	\beq
	\sum_n = \sum_n \frac{\Delta \mu^2_n}{\Delta \mu^2_n} \,\to\,\int_{m_0^2}^\infty \,d\mu^2 \,(R/2) \,\frac{1}{\sqrt{\mu^2-m_0^2}} =  \int_{m_0^2}^\infty \,d\mu^2 \, R \rho(\mu^2)\,,
	\eeq{sum_limit}
where $\rho$ is the spectral density function given in Eq.~\leqn{rho}. Plugging into Eq.~\leqn{PropKK1} 
	\beq
\Pi(p^2, 0) =  \int_{m_0^2}^\infty \,\frac{d\mu^2}{2\pi} \,\frac{i}{p^2-\mu^2+i\epsilon}\, \rho(\mu^2).
\eeq{PropKK2}
 This is exactly the same spectral-density representation of the brane-to-brane propagator as in Eq.~\leqn{SD}. This derivation makes it explicit that the physical meaning of the spectral density $\rho(\mu^2)$ is (up to an overall constant) the density of KK states with respect to $\mu^2$.  

\subsection{Hilbert Space, Orthonormality and Completeness}

The one-particle Hilbert space of our theory in infinite 5D is spanned by basis states $|{\bf p}, k\rangle$. These are eigenstates of 5D momentum, with eigenvalues given by $(E_{\bf p, k}, {\bf p}, k)$ where $E_{\bf p, k}=\sqrt{{\bf p}^2+k^2+m_0^2}$. These states obey orthonormality and completeness relations:
\beqa
& & \int \frac{d^3 p}{(2\pi)^3}\,\frac{dk}{2\pi}\,\frac{1}{2E_{\bf p, k}}|{\bf p}, k\rangle \langle {\bf p}, k| = 1,\CR
& & \langle {\bf p}^\prime, k^\prime|{\bf p}, k\rangle = (2\pi)^4\,(2E_{\bf p, k})\,\delta^3({\bf p}^\prime-{\bf p})\,\delta(k^\prime-k).
\eeqa{comp}    	
The factors of energy are a matter of convention; with our choice the scalar product of basis states is Lorentz-invariant.

Now, let us define $\mu^2\equiv k^2+m_0^2$. We can choose to label our basis states by $\mu^2$ instead of $k$: $|{\bf p}, \mu^2 \rangle \equiv |{\bf p}, k=\sqrt{\mu^2-m_0^2} \rangle$. Note that
\beq
\frac{d\mu^2}{dk} = 2 k = 2 \sqrt{\mu^2-m_0^2} = \frac{1}{\rho(\mu^2)}.
\eeq{varchange}	
Using this Jacobean, the orthonormality and completeness relations become
\beqa
& & \int \frac{d^3 p}{(2\pi)^3}\,\frac{d\mu^2}{2\pi}\,\frac{1}{2E_{\bf p, \mu^2}}\,\rho(\mu^2)\,|{\bf p}, \mu^2\rangle \langle {\bf p}, \mu^2| = 1,\CR
& & \langle {\bf p}^\prime, \mu^{\prime 2}|{\bf p}, \mu^2 \rangle = (2\pi)^4\,\frac{2E_{{\bf p}, \mu^2}}{\rho(\mu^2)}\,\delta^3({\bf p}^\prime-{\bf p})\,\delta(\mu^{\prime 2}-\mu^2)\,,
\eeqa{eq:comp1_app}    	
where $E_{{\bf p}, \mu^2}=\sqrt{{\bf p}^2+\mu^2}$. This is precisely the formulas given in the main text (see \cref{compl1} and (\cref{normal})).   	

An alternative derivation of this result is to start with a compactified 5D space, where the one-particle states are the usual KK modes $|{\bf p}, n\rangle$. These obey the orthonormality and completeness relations 	
\beqa
& & \sum_n  \int \frac{d^3 p}{(2\pi)^3}\,\frac{1}{2E_{{\bf p}, n}}|{\bf p}, n\rangle \langle {\bf p}, n| = 1,\CR
& & \langle {\bf p}^\prime, n^\prime|{\bf p}, n\rangle = (2\pi)^3\,(2E_{{\bf p}, n})\,\delta_{n n^\prime}\,\delta^3({\bf p}^\prime-{\bf p})\,.
\eeqa{compKK}    	
Here $E_{{\bf p}, n} = \sqrt{{\bf p}^2+m_n^2}$. 
The continuum limit in the completeness relation is obtained by replacing 
\beq
\sum_n \to R \int d\mu^2 \rho(\mu^2), 
\eeq{sum_lim}
as in Eq.~\leqn{sum_limit} above. In the orthonormality relation, this continuum limit is taken using
\beq
\delta_{n, n^\prime} \to \frac{1}{R\,\rho(\mu^2)}\,\delta(\mu^{\prime 2}-\mu^2). 
\eeq{delta_cont} 
Rescaling the one-particle states to define 
\beq
|{\bf p}, \mu^2\rangle \equiv \lim_{R\to \infty} \sqrt{R}\, \left\vert {\bf p}, n=R\sqrt{\mu^2-\mu_0^2} \right\rangle\,, 
\eeq{state_lim}
we again reproduce the orthonormality and completeness relations used in the main text.  

\subsection{Boltzmann Equation}

The 5D flat-space model also gives a useful illustration of Boltzmann equations for gapped continuum. A gas of single-particle excitations in this model can be described by a 5D phase-space distribution $f({\bf p}, k)$. To add interactions, let us consider a toy model where the 5D field $\Phi$ is coupled to a 4D field $h(x)$ localized on a brane at $z=0$, via
\beq
S_{\rm int} = \int d^4 x \frac{\lambda}{4} \Phi^2 h^2.  
\eeq{intaraction}     
This interaction enables a $2\to 2$ scattering processes $\Phi\Phi \leftrightarrow h h$. Notice that 4D momentum is conserved in this scattering, but 5D momentum is not, due to the localized nature of the interaction. The standard textbook derivation of the Boltzmann equation trivially generalizes to the flat 5D space, yielding\footnote{This form of the Boltzmann equation would apply for any interaction between two 5D fields and two fields localized on a 4D delta-function brane. For example, $h$ can be replaced by a particle with spin.}
\beqa
E \,\frac{\partial f({\bf p}, k, t)}{\partial t} &=& - \frac{1}{2} \,\int d\Pi^{\prime (5)} \, d\Pi_A d\Pi_B \,
(2\pi)^4 \delta^4 (q_A+q_B-p-p^\prime)\, \CR\CR & &\times |{\cal M}|^2\,\left( f f^\prime (1\pm f_A) (1\pm f_B) - f_A f_B (1\pm f)(1\pm f^\prime)\right).
\eeqa{BE_5D}  
In the collision term on the right-hand side, $q_A$ and $q_B$ denote the 4-momenta of the $h$ particles in the collision, while $P=(p^\mu, k)$ and $P^\prime=(p^{\prime\mu}, k^\prime)$ are the 5-momenta of the $\Phi$ particles. Once again, only the 4-momentum is conserved, as reflected in the delta function in the collision term. The LIPS volume elements for the 4D $h$ particles take their usual form, $d\Pi_{A,B}\equiv \frac{d^3 q_{A,B}}{(2\pi)^3}\,\frac{1}{2E_{A,B}}$, while the LIPS volume element for the 5D $\Phi$ particle with momentum $P^\prime$ has the form 
\beq
d\Pi^{\prime (5)} \equiv \frac{dk^\prime}{2\pi}\,\frac{d^3 p^\prime}{(2\pi)^3}\,\frac{1}{2E^\prime}\,=\,\frac{dk^\prime}{2\pi}\,d\Pi_{\mu^\prime} ,
\eeq{PSel}
where $E^\prime = \sqrt{{\bf p}^{\prime 2}+k^{\prime 2}+m_0^2}$ is the particle's energy, and in the second equality $d\Pi_{\mu^\prime}$ is the usual LIPS volume element for a 4D particle with 3-momentum ${\bf p}^\prime$ and mass $\mu^\prime = \sqrt{k^{\prime 2}+m_0^2}$. Changing the integration variable from $k^\prime$ to $\mu^{\prime 2}$ yields 
\beqa
E \, \frac{\partial f({\bf p}, k, t)}{\partial t} &=& - \frac{1}{2} \,\int \frac{d\mu^{\prime 2}}{2\pi}\,\rho(\mu^{\prime 2})\,\int d\Pi_{\mu^\prime}\,
d\Pi_A d\Pi_B \,
(2\pi)^4 \delta^4 (q_A+q_B-p-p^\prime)\, \CR\CR & &\times |{\cal M}|^2\,\left( f f^\prime (1\pm f_A) (1\pm f_B) - f_A f_B (1\pm f)(1\pm f^\prime)\right),
\eeqa{BE_5D_final}  
where once again the spectral density $\rho$ arises as the Jacobean of the variable change. This is precisely Eq.~\leqn{BE} which formed the basis of our discussion of non-equilibrium thermodynamics in \cref{sec:physics_gapped_continuum} and \cref{sec:freeze-out_DM}.

\section{Spectral Density Near the Gap}
\label{appendix:spectral_density_near_gap}

As we discussed in \cref{sec:Intro} and \cref{sec:preview}, most of the continuum DM phenomenology is governed by the shape of the spectral density near the gap scale.
In \cref{sec:DM_spectral_density_5D} we presented the expression for the spectral density near the mass gap \cref{eq:rho_near_gap} for the particular background geometry of \cref{eq:A(y)} and \cref{eq:phi(y)}. Here we would like to argue that the characteristic square root form obtained is quite general, and applicable to any case with a gapped continuum described by the Schr{\"o}dinger equation
\beq
-\frac{d^2\psi}{dz^2}+V(z)\psi = \kappa^2 \psi.
\eeq{sch}
were $\kappa^2=p^2-\mu_0^2$ is the distance from the gap while as before $\mu_0$ is the gap scale. We assume that $V(z)$ is positive-definite and $V\to 0$ as $z\to+\infty$ (the constant corresponding to the gap is already included in the definition of $\kappa^2$). This equation is guaranteed to have two {\it real}, linearly-independent solutions $\psi_1$ and $\psi_2$.  The general solution (up to an irrelevant overall constant) is\footnote{In fact, the overall factor is in general a function of $p^2$. This however cancels out in $\Sigma (p)$, and hence in $\rho (p^2)$.}
\beq
\psi = \psi_1 + c \psi_2,
\eeq{sol} 
where $c$ can be complex. At large $z$, $V$ can be ignored and the equation can be solved: 
\beq
\psi_1=\cos \kappa z,~~\psi_2=\sin \kappa z. 
\eeq{largez_sol}
The outgoing wave boundary condition at $z\to +\infty$ is $\psi \sim e^{+i\kappa z}$, which fixes $c=i$. 

We recall that the spectral density is given by
\bea
\rho (p^2) = - 2 {\rm Im} \frac{1}{\Sigma (p)} = -2 \frac{{\rm Im} \bar{\Sigma} (p)}{\vert \Sigma (p) \vert^2}
\eea
where $\Sigma (p) = k \partial_z \left( e^{\frac{3}{2} A(z)} \frac{\psi (z,p)}{\psi(R,p)} \right)_{z=R}$\footnote{In \cref{sec:DM_spectral_density_5D} we denoted the profile as $f(z,p)$, which satisfies the same Schr{\"o}dinger equation.} and $\bar{\Sigma}$ is the complex conjugate of $\Sigma$. Obviously, ${\rm Im} \bar{\Sigma} = - {\rm Im} \Sigma$. Explicit computation shows that
\begin{equation}
\rho (p^2) = 2 \frac{{\rm Im} \Sigma (p)}{k} \left[ \left( \frac{3}{2} \left( k + 1/y_s \right) + \frac{{\rm Re} (\bar{\psi} \psi')}{|\psi|^2} \right)_{z=R}^2 + \left( {\rm Im} \Sigma \right)^2 \right]^{-1}
\label{eq:rho_explicit_app}
\end{equation}
where we remind that $k$ is AdS curvature scale (while $\kappa = \sqrt{p^2 - \mu_0^2}$). The point of this expression is that the behavior of $\rho (p^2)$ at the gap scale $p^2 \to \mu_0^2$ is understood from that of ${\rm Im} \Sigma$. Also, relatedly, the regularity (or singularity) is determined by how $\vert \psi \vert^2$ behaves at $z=R$, the UV-brane scale. For this reason, from now on, we focus on (note that $A(z)$ is a real function)
\bea
{\rm Im} \Sigma (p) = \left. k \; {\rm Im} \; \frac{d}{dz} \log \psi \right\vert_{z=R}.
\eea

The log-derivative of the wavefunction is given by 
\beq
\frac{d}{dz} \log \psi = \frac{\psi_1^\prime + i \psi_2^\prime}{\psi_1 + i \psi_2}
\eeq{logdir}
and its imaginary part (remembering that both $\psi_i$'s are real) is
\beq
{\rm Im}~\frac{d}{dz} \log \psi = \frac{\psi_1 \psi_2^\prime - \psi_1^\prime \psi_2}{|\psi_1|^2 + |\psi_2|^2}.
\eeq{im}
The numerator is the Wronskian $W$. Since Eq.~\leqn{sch} has no first-derivative term, by the Abel identity $W$ is a $z$-independent constant. We can compute $W$ at large $z$ using Eq.~\leqn{largez_sol}: $W=\kappa\,(\cos^2 \kappa z + \sin^2 \kappa z) = \kappa$. So we have
\beq
{\rm Im}~\frac{d}{dz} \log \psi = \frac{\kappa}{|\psi(z)|^2} = \frac{\sqrt{p^2-\mu_0^2}}{|\psi(z)|^2}.
\eeq{im2}
This is valid at any $z$, in particular at the location of the UV brane. Therefore 
\beq
\rho (p^2) \propto \frac{\sqrt{p^2-\mu_0^2}}{|\psi(R)|^2}.
\eeq{SD}
This almost proves that $\rho (p^2) \to 0$ as $p^2\to \mu_0^2$; the only caveat is that we still need to prove that $|\psi(R)|\not=0$ in this limit. A simple argument is that since $z=R$ is arbitrary, and we cannot have $|\psi(R)|=0$ for more than a finite set of values of $z=R$, generically we should expect it to be non-zero. However a stronger argument can be constructed that in fact it {\it cannot} be 0. Multiply Eq.~\leqn{sch} on both sides by $\psi^*$, and integrate over $z$ from some $z_0$ to $+\infty$. Then use integration by parts on the first term. This gives
\beq
\left. -\left(\frac{d}{dz}|\psi|^2 \right)\right\vert^{+\infty}_{z_0} \,+\, \int_{z_0}^{+\infty} dz\,\left[ |\psi^\prime|^2 + V(z) |\psi|^2\right] = \kappa^2 \int_{z_0}^{+\infty} dz\, |\psi|^2.
\eeq{integral} 	
Dropping terms proportional to powers of $\kappa$ (as $\kappa \to 0$ for modes $p^2 \to \mu_0^2$), we have
\beq
\frac{d}{dz}|\psi|^2 (z_0)	+ \int_{z_0}^{+\infty} dz\,\left[ |\psi^\prime|^2 + V(z) |\psi|^2\right] = 0
\eeq{integral1}	
which implies 
\beq
\frac{d}{dz}|\psi|^2 (z_0) < 0,
\eeq{ineq}
for modes very close to the gap scale $\mu_0$.
Since $z_0$ was arbitrary, this means that $|\psi|^2$ is a monotonically decreasing function, and since we know that  $|\psi|^2=1$ at large positive $z$ (see eq.~\leqn{largez_sol}), it means that $|\psi|^2\geq 1$ for any $z$ and therefore non-zero at the UV-brane $z = R$. 

One may still worry that the wave function itself is diverging at $z=R$, and hence strongly influencing the way the spectral density goes to zero at the gap. However, again $z=R$ is not a special point in the geometry, but in fact arbitrarily chosen, hence the potential or wave function is not expected to have a singularity at $z=R$. Therefore for the generic case, we expect that $\lim_{\kappa\to 0} |\psi(R)|^2 = C$, some \emph{finite} constant. In this case according to eq.~\leqn{SD}, 
\bea
\rho (p^2) \propto \sqrt{p^2-\mu_0^2}.
\eea
One implication of this is that $\rho (p^2) \to 0$ as $p^2 \to \mu_0^2$.
Notice also that as long as $\lim_{\kappa\to 0} |\psi(R)|^2 = C < \infty$, $\rho (p^2)$ is regular for $p^2$ close to $\mu_0^2$ as is seen directly from \cref{eq:rho_explicit_app}.
In fact, this is exactly what we have found in \cref{sec:DM_spectral_density_5D} computed in the background, \cref{eq:A(y)} and \cref{eq:phi(y)}, both using the analytic asymptotic solution as well as the numerical solution.  
It may be worth mentioning that it is possible to solve the \cref{sch} analytically for $V(z) \propto e^{-a z}$ for any constant $a$ and one finds that ${\rm Im} \Sigma \propto \sqrt{p^2 - \mu_0^2}$. In \cref{sec:DM_spectral_density_5D}, we indeed have shown that the form of the potential at $z \to \infty$ do have this form with $a = \frac{2}{3} \mu_0$.

\section{AdS/CFT Duality of Gapped Continuum}
\label{app:AdS_CFT}

In this appendix, we describe 4D CFT dual description of the 5D $Z$-portal model introduced in \cref{subsec:5D_Z_portal}.
The boundary effective action $S_{\rm eff} [\hat{\Phi}]$ derived in \cref{subsec:5D_Z_portal} is interpreted as (the leading order in large-$N$ expansion) partition function of the dual CFT via the AdS/CFT correspondence \cite{Maldacena:1997re, Witten:1998qj, Gubser:1998bc}:
\bea
Z_{\scaleto{\rm AdS}{4pt}} [\hat{\Phi}] &=& \int_{\scaleto{\rm AdS}{4pt}} \mathcal{D} \Phi \vert_{\Phi \vert_{\scaleto{\rm UV}{3pt}} = \hat{\Phi}} \; e^{i S_{\scaleto{\rm AdS}{4pt}} [\Phi]} \approx e^{i S_{\rm eff}} [\hat{\Phi}] \\
&=& \int_{\scaleto{\rm CFT}{4pt}} \; \mathcal{D} \varphi \; e^{i S_{\scaleto{\rm CFT}{4pt}} [\varphi] + i S_{\rm ext} [\hat{\Phi}] + i \int \frac{1}{\Lambda^{d-3}} \hat{\Phi}^\dagger \mathcal{O} + {\rm h.c.}} = Z_{\scaleto{\rm CFT}{4pt}} [\hat{\Phi}].
\eea
where $d$ is the scaling dimension of the CFT operator $\mathcal{O}$ sourced by the UV boundary value $\hat{\Phi}$ of 5D field $\Phi$.
With finite UV cutoff scale $\Lambda$, the source field $\hat{\Phi}$ may be dynamical, and for this reason we added the action $S_{\rm ext} [\hat{\Phi}]$ for the external field $\hat{\Phi}$. 
From this, we can derive the relation between CFT two point function and $\Sigma (p^2)$
\begin{equation}
\Sigma (p^2) = \frac{i}{\Lambda^{2d-6}} \langle \mathcal{O} \mathcal{O}^\dagger \rangle (p) + G_{\hat{\Phi}} (p^2)
\label{eq:CFT_2_pt_function_and_Sigma}
\end{equation}
where $G_{\hat{\Phi}} (p^2)$ is two point function of $\hat{\Phi}$ obtained from $S_{\rm ext} [\hat{\Phi}]$, possibly including wave function renormalization factor. For instance, if $S_{\rm ext} [\hat{\Phi}] = Z \partial_\mu \hat{\Phi}^\dagger \partial^\mu \hat{\Phi}$ then $G_{\hat{\Phi}} (p^2) = Z p^2$. In the IR, the dual 4D QFT goes through a phase transition into gapped continuum phase. In this phase, a CFT operator creates composite (gapped) continuum states and the source term $\hat{\Phi}^\dagger \mathcal{O}$ describes a mixing between external degree of freedom $\hat{\Phi}$ and composite continuum modes. To understand this mixing better, we may write the CFT operator in terms of canonically normalized field $\phi_\mu (x)$ which excites a mode with $p^2 = \mu^2$ as
\begin{equation}
\mathcal{O} (x) = \mu_0^{d-1} \int_1^\infty  \frac{d (\mu /\mu_0)}{2\pi} \; c \left( \mu / \mu_0 \right) \phi_\mu (x).
\label{eq:CFT_O_in_terms_of_modes}
\end{equation}
Here, $\mu_0$ is the gap scale and the dimensionless function $c (\mu / \mu_0)$ has a support from $\mu_0$ to some $\mu \sim \mathcal{O} (\Lambda)$, hence determines the integration upper limit. As we show below, this function is directly related to the spectral density. Using 
\bea
\langle \phi_\mu^\dagger (p) \phi_{\mu'} (k) \rangle = \frac{i}{p^2 - \mu^2 + i \epsilon} (2\pi)^4 \delta^4 (p-k) (2\pi) \delta \left( \frac{\mu}{\mu_0} - \frac{\mu'}{\mu_0} \right)
\eea 
we can rewrite the continuum part in \cref{eq:CFT_2_pt_function_and_Sigma} as
\bea
\Sigma_c (p^2) \equiv \Sigma (p^2) -  G_{\hat{\Phi}} (p^2) = \frac{\mu_0^{2d-3}}{\Lambda^{2d-6}} \int_{\mu_0^2}^\infty \frac{d \mu^2}{2\pi} \; \frac{\left[ c \left( \mu / \mu_0 \right)^2 / 2 \mu \right]}{\mu^2 - p^2 - i \epsilon} .
\eea
This in turn implies that
\begin{equation}
{\rm Im} \Sigma_c (p^2) =  \frac{\mu_0^{2d-3}}{\Lambda^{2d-6}} \left[ c \left( p / \mu_0 \right)^2 / 4 p \right], \;\;\;\; p = \sqrt{p^2}.
\label{eq:Sigma_and_c}
\end{equation}
In addition, the mixing between the external field and the composite continuum modes is readily found to be
\begin{equation}
\mathcal{L} \supset \frac{1}{\Lambda^{d-3}} \hat{\Phi}^\dagger \mathcal{O}  =  \int \frac{d (\mu / \mu_0)}{2\pi} \; \sqrt{4\mu_0 \mu {\rm Im} \Sigma_c} \; \hat{\Phi}^\dagger \phi_\mu 
\label{eq:CFT_mixing_in_terms_of_ImSigma}
\end{equation}
where we used \cref{eq:Sigma_and_c} to get the final expression.
\begin{figure}
\begin{center}
\includegraphics[width=13cm]{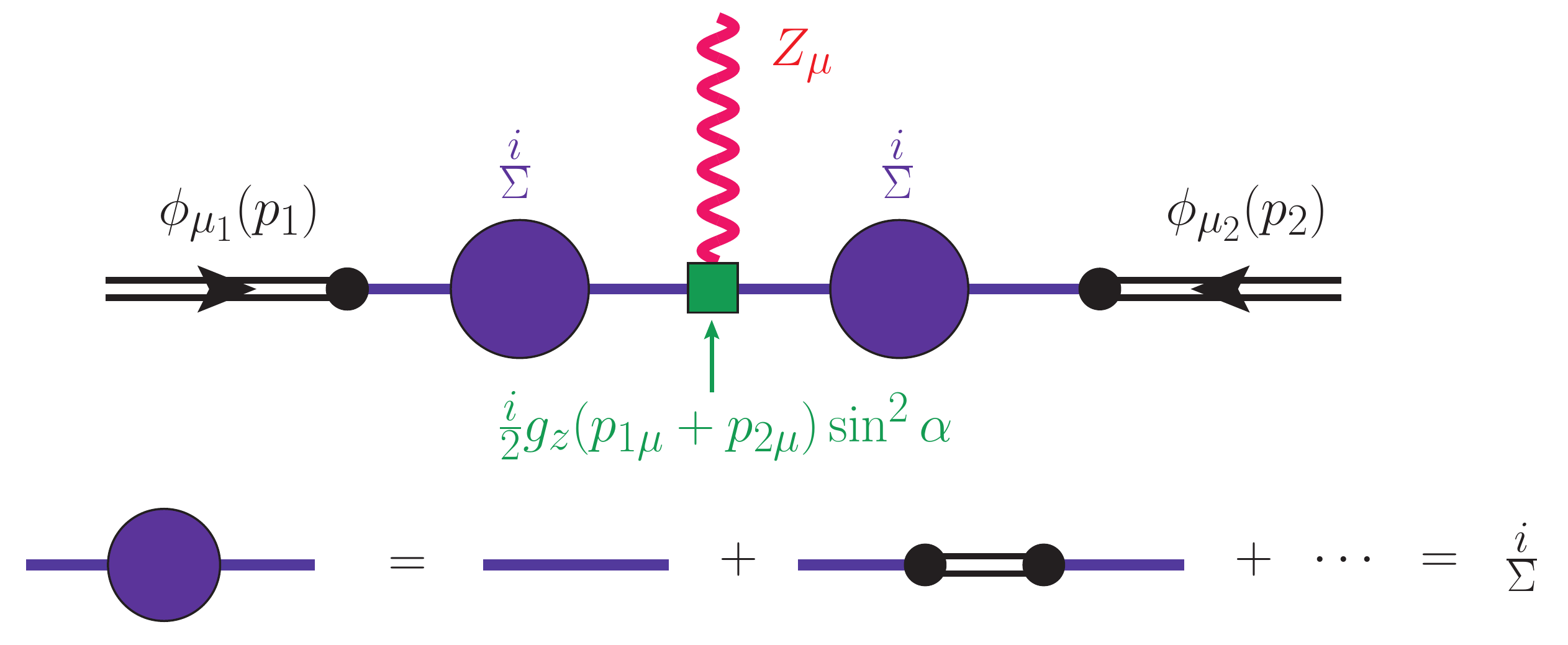} 
\end{center}
\caption{Effective coupling in CFT picture through the resummation of $\hat{\Phi} {\rm -} \phi_\mu$ mixing. Composite continuum modes $\phi_\mu$ are denoted as double lines and the external field $\hat{\Phi}$ is depicted with a single line. }
\label{fig:AdS_CFT_resum} 
\end{figure}

In the CFT picture, it is the external field $\hat{\Phi}$ that couples to the SM $Z$ and $W$. Then the coupling of continuum modes to the $Z$ and $W$ is obtained through mixing. One thing that we need to be a bit careful is that we need to resum all the diagrams to get reliable answer, since the $\mu$-dependent mixing given in \cref{eq:CFT_mixing_in_terms_of_ImSigma} is generally not small. This may be done by first computing resummed $\hat{\Phi}$-propagator and inserting such resummed propagator into the relevant Feynman diagrams. It is straightforward to show that the resummation of the diagrams in \cref{fig:AdS_CFT_resum} yields
\bea
\langle \hat{\Phi} \hat{\Phi}^\dagger \rangle (p) = \frac{i}{G_{\hat{\Phi}}} + \frac{i}{G_{\hat{\Phi}}} \left( \frac{i}{\Lambda^{d-3}} \right)^2 \langle \mathcal{O} \mathcal{O}^\dagger \rangle \frac{i}{G_{\hat{\Phi}}} + \cdots = \frac{i}{G_{\hat{\Phi}} + i \frac{1}{\Lambda^{2d-6}} \langle \mathcal{O} \mathcal{O}^\dagger \rangle} = \frac{i}{\Sigma (p^2)}.
\eea
In fact, we could have obtained this easily from holographic effective action by viewing it as an action for $\hat{\Phi}$ including CFT contributions (i.e.~resummation).
Then the coupling of a pair of continuum modes ($\mu_1$ and $\mu_2$) to the $Z$ boson is given by
\bea
g_{\rm eff} = \left[ \frac{i}{2} g  \left( p_1 + p_2 \right)_\mu \right] \left[  \sin \alpha_{\mu_1} \sqrt{ 2 \mu_0 \mu_1 \rho (\mu_1^2)}  \right] \left[ \sin \alpha_{\mu_2} \sqrt{  2 \mu_0 \mu_2 \rho (\mu_2^2)} \right].
\eea
To get this, we used 
\bea
\rho (\mu^2) = -2 {\rm Im} \Sigma^{-1} = 2 \frac{{\rm Im} \Sigma_c}{\vert \Sigma \vert^2}
\eea
and \cref{eq:CFT_mixing_in_terms_of_ImSigma}. Note that ${\rm Im} \Sigma = {\rm Im} \Sigma_c$. We also remind that the first factor is the usual coupling to $Z$-boson, and $\sin \alpha_\mu$ is from the mixing $\hat{\Phi}$ with $\chi$ that directly couples to $Z$ (see \cref{sec:Z_portal_model}). Therefore, we see that the effective coupling is a product of the usual $Z$-coupling of a complex scalar (the first factor) and mixing angle for each continuum modes (the second and third factors).

Finally, let us compute the cross section for a process in which two SM particles $A$ and $B$ annihilate into a pair of continuum DM with $\mu_1$ and $\mu_2$ through the $Z$-exchange. For concreteness, let us assume (as we did so far) that $\sin \alpha_\mu$ is $\mu$-independent, which may be achieved by taking $m_\chi \gg \mu$. Denoting $\delta_\mu \equiv \sqrt{ 2 \mu_0 \mu \rho (\mu^2)}$, the matrix amplitude may be written as
\bea
\mathcal{M} \left( {\rm A} + {\rm B} \to {\rm DM (\mu_1)} + {\rm DM (\mu_2)} \right) = \delta_{\mu_1} \delta_{\mu_2} \hat{\cal M}
\eea
where we factored out the entire $\mu$-dependent piece in the form of mixing angles, and defined the $\mu$-independent matrix element $\hat{\cal M}$ which is just a matrix element for particles with mass $\mu_1$ and $\mu_2$. The cross section with the possible final states summed over (at the level of rate, not the matrix element) then is given by
\bea
\sigma \left( {\rm A} + {\rm B} \to {\rm DM (\mu_1)} + {\rm DM (\mu_2)} \right) = \int \frac{d \mu_1^2}{2\pi} \; \rho (\mu_1^2) \int \frac{d \mu_2^2}{2\pi} \; \rho (\mu_2^2) \; \hat{\sigma}
\eea
where similarly to the matrix element, $\hat{\sigma}$ is a cross section computed in terms of $\hat{\cal M}$ in a way that continuum modes are treated as particles with mass $\mu_1^2$ and $\mu_2^2$. Clearly then, the phase space density for these canonically normalized continuum modes are the usual Lorentz invariant measure given in \cref{Pies}. Crucially, this final expression, obtained with mixing angles and proper mode-sum defined through \cref{eq:CFT_O_in_terms_of_modes}, agrees exactly with \cref{eq:xsec}.

\bibliographystyle{utphys}
\bibliography{ref}

\end{document}